\documentclass[aps,nofootinbib,amsmath,prd,twocolumn,notitlepage,showpacs,superscriptaddress]{revtex4-1}

\usepackage[american]{babel}
\usepackage{amsfonts}
\usepackage{amsmath}
\usepackage{amssymb}
\usepackage{epsfig}
\usepackage{latexsym}
\usepackage{paralist}
\usepackage{graphicx}
\usepackage[vcentermath]{youngtab}
\usepackage{young}
\usepackage{ytableau}
\usepackage{etex}
\usepackage{braket}
\usepackage{float}
\usepackage{slashed}
\usepackage{xcolor}
\usepackage{soul}
\usepackage{dcolumn}
\usepackage{bm}
\usepackage{nicefrac}
\usepackage{tipa}
\usepackage{dsfont}

\begin{document}



%
%

\title{PROJECTIVE TRANSFORMATIONS IN METRIC-AFFINE AND WEYLIAN GEOMETRIES}

\author{DARIO SAURO}
\address{
Universit\`a di Pisa, Largo Bruno Pontecorvo 3\\
 Pisa, 56127, Italy}
\address{INFN - Sezione di Pisa, Largo Bruno Pontecorvo 3\\
Pisa, 56127, Italy\\
dario.sauro@phd.unipi.it}

\author{RICCARDO MARTINI}
\address{
INFN - Sezione di Pisa, Largo Bruno Pontecorvo 3\\
Pisa, 56127, Italy\\
riccardo.martini@pi.infn.it}

\author{OMAR ZANUSSO}
\address{
Universit\`a di Pisa, Largo Bruno Pontecorvo 3\\
Pisa,  56127, Italy}
\address{INFN - Sezione di Pisa, Largo Bruno Pontecorvo 3\\
Pisa, 56127, Italy\\
omar.zanusso@unipi.it}



\begin{abstract}
%
We discuss generalizations of the notions of projective transformations
acting on affine model of Riemann-Cartan and Riemann-Cartan-Weyl gravity
which preserve the projective structure of the light-cones.
We show how the invariance under some projective transformations can be used to
recast a Riemann-Cartan-Weyl geometry either as a model in which the role of the Weyl gauge potential is played by the torsion vector, which we call torsion-gauging,
or as a model with traditional Weyl (conformal) invariance.
\end{abstract}

\maketitle

\renewcommand{\thefootnote}{\arabic{footnote}}
\setcounter{footnote}{0}

\section{Introduction}\label{sect:intro}

Projective invariances are transformations that preserve the so-called projective structure, defined as the set of all solutions of the autoparallel equation of a manifold~\cite{Schouten2013ricci,kobayashi1964projective,ehlers1973geometry,aminova1995projective,giachetta1997projective,matveev2018projectively}.
Since the autoparallel equation depends only on a connection, projective-like transformations are becoming increasingly important in the context of metric-affine gravity, because metric and connection can be transformed independently~\cite{Afonso:2017bxr,Aoki:2019rvi,BeltranJimenez:2019acz,BeltranJimenez:2020sqf,BeltranJimenez:2020guo,Klemm:2020mfp,Garcia-Parrado:2020lpt,Iosifidis:2018zwo,Iosifidis:2018zij,Percacci:2020ddy}.
As a consequence, the invariance under some generalized projective transformations
could be used as a symmetry requirement, together with diffeomorphism and local Lorentz
invariance, to construct sensible gravitational theories.
This becomes particularly relevant for general metric-affine theories,
because there is a proliferation of possible interaction terms given that
the independent connection allows for many possible contractions, even for
low-order vertices \cite{Baldazzi:2021kaf}.
Projective invariances have also recently received renewed attention thanks to their relevance when preventing ghost-like instabilities \cite{BeltranJimenez:2019acz,Aoki:2019rvi,BeltranJimenez:2020sqf,Percacci:2020ddy}.

In this paper we follow the lead
of Baldazzi et al.~\cite{Baldazzi:2021kaf} (see also Iosifidis~\cite{Iosifidis:2019fsh})
and begin by discussing a generalized set of projective transformations
that are vector-like (in the sense that they are generated by vectors in four dimensions).
We show what are the implications on the main components of the connection, which include the contortion and the disformation tensors.

Then, in the most important part of the paper, we specialize the application of the generalized projective transformations to a special type of metric-affine theory, that is, to a Riemann-Cartan-Weyl theory in which local dilatations are gauged through an
additional Abelian vector potential (the Weyl potential) \cite{Charap:1973fi,Gasperini:2017ggf,Ghilencea:2018dqd,Ghilencea:2018thl,Ghilencea:2019jux}.
The main result is that a general Riemann-Cartan-Weyl theory
that is constrained through the use of a combination of projective symmetries
can be interpreted as a theory in which the vector component of the torsion
plays the role of the Weyl potential, recently put to use in a cosmological context by Karananas et al.~\cite{Karananas:2021gco}. We refer to this type of model as ``torsion-gauged'' (see also Ref.~\cite{Karananas:2015eha} for more details on this procedure).

An interesting aspect of our construction is that the truly tensorial part of the independent connection is a Weyl-invariant Curtright-like tensor with mixed symmetries \cite{Curtright:1980yk},
whose interactions are, for the most part, not affected by the projective transformations, so it is a natural tensor to include in extensions of
the torsion-gauged cosmological models with Weyl invariance.
This is shown explicitly through some examples in the paper.

\section{Metric-affine theories}\label{sect:intro-mags}

Consider a $d$-dimensional spacetime manifold equipped with a metric $g_{\mu\nu}$ of Lorenzian signature and an independent connection $\nabla$, which acts on vectors as
$\nabla_\mu v^\nu = \partial_\mu v^\nu + \Gamma^\nu{}_{\rho\mu} v^\rho$, where $\Gamma$
are the holonomic components of the connection.
The traditional formulation of general relativity (GR) is based on giving a dynamical origin to the metric through Einstein's equations and adopts the unique metric-compatible and symmetric Levi-Civita connection \cite{Weinberg:1972kfs,Wald:1984rg}, which we denote $\mathring{\nabla}$ with components $\mathring{\Gamma}$. The resulting geometry is a pseudo-Riemannian manifold.

In metric-affine theories of gravity (MAGs), metric and connection are regarded as independent entities and the dynamics of the system must be supplemented by additional equations. \cite{Hehl:1976my,Percacci:1990wy,Hehl:1994ue,Percacci:2009ij,BeltranJimenez:2019esp}
Nevertheless, the difference of two connections is a tensor, from which we deduce that it is always possible to express any independent connection as the sum of the Levi-Civita one and a tensor, as $\Gamma^\nu{}_{\rho\mu}
=\mathring{\Gamma}^\nu{}_{\rho\mu}  +\Phi^\nu{}_{\rho\mu} $. It is not difficult to prove that
\begin{equation}\label{eq:Gamma}
\Gamma^\nu{}_{\rho\mu} = \mathring{\Gamma}^\nu{}_{\rho\mu} + K^\nu{}_{\rho\mu} + N^\nu{}_{\rho\mu}\, ,
\end{equation}
where $K$ is known as the contortion tensor and depends only on the antisymmetric part of $\Gamma$ given by the torsion tensor, $T^\nu{}_{\rho\mu}=\Gamma^\nu{}_{\mu\rho} - \Gamma^\nu{}_{\rho\mu}$
\begin{equation}\label{eq:Contortion}
K_{\nu\rho\mu} =  \frac{1}{2} \left( T_{\rho\nu\mu} + T_{\mu\nu\rho} - T_{\nu\rho\mu} \right) \,,
\end{equation}
while $N$ is known as the distortion tensor and depends only on the nonmetricity tensor, $Q_{\mu\nu\rho} = - \nabla_\mu g_{\nu\rho}$,
\begin{equation}\label{eq:Distortion}
 N_{\rho\nu\mu} = \frac{1}{2} \left( Q_{\mu\nu\rho} + Q_{\nu\mu\rho} - Q_{\rho\mu\nu} \right) \, .
\end{equation}
The connection \eqref{eq:Gamma} is a general $GL(d)$ connection for spacetime, and it is straightforward to prove that, if we require it to be symmetric and metric-compatible, then disformation and contortion are zero, implying $\Gamma= \mathring{\Gamma}$.\footnote{%
Notice that the nonmetricty is often defined with a different sign, i.e., $Q=+\nabla g$.
Furthermore, the symbol $L$ is often used for the disformation in place of our unconventional choice $N$,
but we reserve $L$ for a more special contribution which appears later.
In order to prove \eqref{eq:Gamma}, it is sufficient to take the expression $ \nabla_\mu g_{\nu\rho}+Q_{\mu\nu\rho}=0$, sum and subtract it with itself with cycled indices, then expand the covariant derivatives and collect the terms contracted with the same metrics.
}

Fields of arbitrary Lorentz spin can be coupled to the geometry introducing a coframe $e^a{}_\mu$
such that $g_{\mu\nu}=e^a{}_\mu e^b{}_\nu \eta_{ab}$, which is invariant under local Lorentz transformations $SO(1,d-1) \subset GL(d)$. Using the tetrad postulate, $\nabla e^a=0$, we can determine the spin-connection through its components $\omega^a{}_{b\mu}$ as \cite{Gasperini:2017ggf}
\begin{equation}
\Gamma^\nu{}_{\rho\mu} = E^\nu{}_a \left( \partial_\mu e^a{}_\rho + \omega^a{}_{b\mu} e^b{}_\rho  \right) \, ,
\end{equation}
where $E^\nu{}_a$ is the inverse of $e^a{}_\mu$. Using the spin-connection, the covariant derivative of a Lorentz field $\Psi$ that transforms under the representation with generators $J_{ab}$ of $O(1,d-1)$ is
$\nabla_\mu \Psi = \partial_\mu \Psi + \frac{i}{2}\omega^{ab}{}_\mu J_{ab} \Psi$,
where a summation over the internal indices of the representation is tacitly assumed.
Obviously, this construction can equally well be performed for a general MAG or in the special case of the Levi-Civita connection, in the latter we denote the components as $\mathring{\omega}^{a}{}_{b\mu}$.
It is also straightforward to check the compatibility of the connection when comparing
Lorentz and tangent vectors, i.e., $v^a=e^a{}_\mu v^\mu $, for which it is sufficient to recall the explicit form of the spin-$1$ generators $[J_{ab}]^c{}_d= i (\delta^c_a \eta_{bd}-\delta^c_b \eta_{ad})$.
To elaborate on the last statement, notice that this procedure allows to couple fields with arbitrary Lorentz spin to spacetime, but treats integer and semi-integer spins slightly differently from the point of view of $GL(d)$. While, for example, we can institute a one-to-one correspondence between Lorentz and tangent vectors (as in $v^a=e^a{}_\mu v^\mu $), the same is not true for spinors and other semi-integer representations, whose fields must transform as scalars under general coordinate transformations \cite{Gasperini:2017ggf,Lindwasser:2022nfa}.

\section{Projective structures and transformations}\label{sect:general-projective}

\subsection{Projective structures and invariances}

Consider the motion of a point-particle with velocity $u^\mu=\frac{dx^\mu}{d\tau}$.
The solutions to the auto-parallels equation
\begin{equation}\label{eq:autoparalles}
u^\mu \nabla_\mu u^\nu = f u^\nu
\end{equation}
depends only on the connection $\nabla$ and on an arbitrary scalar function $f(x)$,
besides the initial conditions at some value of the parameter $\tau$. The autoparallels should be considered as the \emph{straightest} lines, as opposed to the shortes lines that arise from minimizing $\int ds$ and obey \eqref{eq:autoparalles} with the Levi-Civita connection. Through a reparametrization of $\tau$ the function $f$ can be eliminated, giving the affinely parametrized auto-parallels equation $u^\mu \nabla_\mu u^\mu=0$.
The images of the solutions are the geodesics associated to $\nabla$. We say that
two connections $\nabla$ and $\nabla'$ belong to the same \emph{projective structure}
if they admit the same geodesics \cite{ehlers1973geometry,aminova1995projective}. Similarly, we say that $\nabla$ and $\nabla'$
belong to the same \emph{lightcone projective structure} if they admit the same lightcone geodesics, i.e., the geodesics for which $u^\mu u_\mu =0$. Notice that the concept of lightcone projective structure is in general independent from the one of conformal structure (generally associated to an equivalence class of conformally related metrics)\cite{Matveev:2020wif}, but it does depend on $g_{\mu\nu}$ through the requirement on the norm of $u^\mu$.

A transformation of the connection that leaves the projective structure invariant is known as projective invariance. If the connection tansforms as $\Gamma^\nu{}_{\rho\mu} \to \Gamma^\nu{}_{\rho\mu} +\delta \Gamma^\nu{}_{\rho\mu}$, which can be understood as a finite transformation, then a projective transformation must satisfy
$\delta \Gamma^\nu{}_{\rho\mu} u^\rho u^\mu= h u^\nu$ for some arbitrary scalar function $h(x)$. In the case of importance for GR, we can think at the Einstein-Hilbert action
as a functional that depends on the metric separately though the direct metric dependence and the connection
\begin{equation}\label{eq:EH-action}
S_{EH}[g_{\mu\nu}] = S_{EH}[g,\mathring{\Gamma}[g]]\,,
\end{equation}
and the resulting functional is invariant under the ``transformation''
\begin{equation}\label{eq:proj-example-1}
\mathring{\Gamma}^\nu{}_{\rho\mu} \to \mathring{\Gamma}^\nu{}_{\rho\mu} + \delta^\nu_\rho j_\mu \,,
\end{equation}
for an arbitrary form $j_\mu$, which also results in a projective invariance \cite{giachetta1997projective,Garcia-Parrado:2020lpt}. This implies that a projective mode is not fixed in MAG theories that use the generalization of $S_{EH}$ to independent connections \cite{Garcia-Parrado:2020lpt} . Notice that we enveloped the word transformation between commas, as, strictly speaking, the components $\mathring{\Gamma}$ do depend only on the metric $g_{\mu\nu}$ that does not transform projectively.

Also of importance for GR, we can now concentrate more on the autoparallel equation. It can be proven that, on a local chart, every projective invariance can be written in the form
\begin{equation}\label{eq:proj-example-2}
\mathring{\Gamma}^\nu{}_{\rho\mu} \to \mathring{\Gamma}^\nu{}_{\rho\mu} + \delta^\nu_\rho v_\mu+ \delta^\nu_\mu v_\rho\,,
\end{equation}
for some vector $v_\mu$ \cite{Schouten2013ricci,thomas:1934differential,aminova1995projective}. In the case of GR, this ``transformation'' should be understood as originating from a more primitive transformation of the metric, i.e., a diffeomorphism.\footnote{%
Ref.~\cite{aminova1995projective} considers two $d$-dimensional Riemannian manifolds $(M,g)$ and $(M',g')$ with corresponding geodesics and such that there exists a diffeomorphism $f:M\rightarrow M'$ such that for every geodesics $\gamma: I \subset \mathbb{R} \rightarrow M$, $f(\gamma)$ is a geodesic in $M'$ and vice-versa (using the inverse diffeomorphism denoted $f^{-1}$). The necessary and sufficient condition for this to happen is that in the corresponding local coordinate systems ($x^\mu |_{p \in M} = x^\mu |_{f(p)\in M'}$) we have $\nabla'_\rho g_{\mu\nu} = 2 g'_{\mu\nu} \partial_\rho \psi + g'_{\mu\rho} \partial_\nu \psi + g'_{\nu\rho} \partial_\mu \psi$, or $\mathring{\Gamma}'^\rho{}_{\nu\mu} - \mathring{\Gamma}^\rho{}_{\nu\mu} = \delta^\rho{}_\nu p_\mu + \delta^\rho{}_\mu p_\nu$, with $p={\rm d} \psi =  {\rm d} ( \frac{1}{2(d+1)} \log (\frac{g}{g'}) )$, where primed covariant derivatives and Christoffel symbols means that they are computed using the primed metric $g'_{\mu\nu}$.
}
Similarly, the transformation
\begin{equation}\label{eq:proj-example-3}
\mathring{\Gamma}^\nu{}_{\rho\mu} \to \mathring{\Gamma}^\nu{}_{\rho\mu} + g_{\rho\mu} w^\nu
\end{equation}
leaves the lightcone projective structure invariant.
It is straightforward to check that \eqref{eq:proj-example-1}
and \eqref{eq:proj-example-2} are such that $\delta \Gamma^\nu{}_{\mu\rho}u^\mu u^\rho \sim u^\nu$ for arbitrary $u^\mu$, while \eqref{eq:proj-example-3} is such that
$\delta \Gamma^\nu{}_{\mu\rho}u^\mu u^\rho =0$ by virtue of $u^\mu$ being lightlike, $u^2=0$.
Furthermore, notice that \eqref{eq:proj-example-2} and \eqref{eq:proj-example-3}
map a symmetric connection into another symmetric connection, as one would expect
in traditional GR applications.

In passing, we notice that a linear combination of \eqref{eq:proj-example-2} and \eqref{eq:proj-example-3} that preserves the Einstein-Hilbert action \eqref{eq:EH-action} up to a boundary term is
\begin{equation}\label{eq:proj-example-4}
\mathring{\Gamma}^\nu{}_{\rho\mu} \to \mathring{\Gamma}^\nu{}_{\rho\mu} + \delta^\nu_\rho v_\mu+ \delta^\nu_\mu v_\rho + \frac{1}{2}c \, g_{\rho\mu} v^\nu\,,
\end{equation}
where the coefficient $c=c(d)$ depends on the dimensionality and has two possible solutions
$ c(d) = d \pm \sqrt{d^2-4}$. Note that in $d=2$ we have only one solution to the above equation.

\subsection{Projective transformations}

Motivated by the above transformations, and following in part \cite{Percacci:2020ddy,Baldazzi:2021kaf},
we define the \emph{generalized projective transformations} by parametrizing the most general vector-based transformations of $\Gamma$ in $d\geq 3$ which make use only of the metric and of the Levi-Civita totally antisymmetric tensor $\varepsilon$
\begin{equation}\label{proj-mag-general}
\begin{split}
\Gamma^\nu{}_{\rho\mu} &\rightarrow  \Gamma^\nu{}_{\rho\mu} + \delta^\nu{}_\rho p_\mu + \delta^\nu{}_\mu q_\rho + g_{\rho\mu} r^\nu
\\&
+ \varepsilon^{\sigma_{1}\cdots \sigma_{d-3}\nu}{}_{\rho\mu} s_{\sigma_{1}\cdots \sigma_{d-3}} \,,
\end{split}
\end{equation}
where $p_\mu$, $q_\mu$ and $r_\mu$ are arbitrary vectors, while $s_{\sigma_{1}\cdots \sigma_{d-3}}$ are the components of an arbitrary $(d-3)$-form. The form $s$
has been included because in the physically interesting case $d=4$ it reduces to an axial vector, and \eqref{proj-mag-general} becomes
\begin{equation}\label{proj-mag-d4}
\Gamma^\nu{}_{\rho\mu} \rightarrow  \Gamma^\nu{}_{\rho\mu} + \delta^\nu{}_\rho p_\mu + \delta^\nu{}_\mu q_\rho + g_{\rho\mu} r^\nu + \varepsilon^{\sigma\nu}{}_{\rho\mu} s_\sigma\,,
\end{equation}
Similarly, in $d=3$, we have that $s$ becomes a pseudo-scalar
\begin{equation}\label{proj-mag-d3}
\Gamma^\nu{}_{\rho\mu} \rightarrow  \Gamma^\nu{}_{\rho\mu} + \delta^\nu{}_\rho p_\mu + \delta^\nu{}_\mu q_\rho + g_{\rho\mu} r^\nu + \varepsilon^{\nu}{}_{\rho\mu} s
\,,
\end{equation}

The generalized projective transformations \eqref{proj-mag-general} induce
a transformation of the tensors introduced in Sect.~\ref{sect:intro-mags}
by keeping $\mathring{\Gamma}$ fixed.
We catalog them here.
For the torsion, we use the definition $T^\nu{}_{\rho\mu}=\Gamma^\nu{}_{\mu\rho} - \Gamma^\nu{}_{\rho\mu}$, from which we deduce
\begin{equation}\label{proj-on-torsion}
\begin{split}
 \delta T^\nu{}_{\rho\mu} & = \delta^\nu{}_\mu (p-q)_\rho - \delta^\nu{}_\rho (p-q)_\mu
 \\&
 + 2 \varepsilon^{\sigma_{1}\cdots \sigma_{d-3}\nu}{}_{\rho\mu} s_{\sigma_{1}\cdots \sigma_{d-3}}\,,
\end{split}
\end{equation}
then, using the relation between torsion and contortion \eqref{eq:Contortion}, we deduce
\begin{equation}\label{proj-on-contortion}
\begin{split}
 \delta K_{\nu\rho\mu} & = g_{\rho\mu} (p - q)_\nu - g_{\nu\mu} ( p - q )_\rho
 \\&
 +\varepsilon^{\sigma_{1}\cdots \sigma_{d-3}}{}_{\nu\rho\mu} s_{\sigma_{1}\cdots \sigma_{d-3}}\,.
\end{split}
\end{equation}
For the nonmetricity, we again use the definition $Q_{\mu\nu\rho} = - \nabla_\mu g_{\nu\rho}$ and the fact that $g_{\mu\nu}$ does not transform to deduce
\begin{equation}\label{proj-on-nonmetricity}
 \delta Q_{\mu\nu\rho} =  2 g_{\nu\rho} p_\mu + g_{\nu\mu} (q+r)_\rho+ g_{\rho\mu} (q+r)_\nu\,,
\end{equation}
which can be used with Eq.~\eqref{eq:Distortion} to find the transformation of the distortion
\begin{equation}\label{proj-on-distortion}
 \delta N_{\nu\rho\mu} =  g_{\nu\rho} p_\mu + g_{\nu\mu} p_\rho + g_{\rho\mu} (q + r - p )_\nu\,.
\end{equation}
Recall that both the distortion and its projective transformation must be symmetric under the exchange of the second and third indices.
The specialization of the above formulas to the physical case $d=4$ is straightforward.

Since all projective transformations are carried out using general $1$-forms, only the ``trace parts'' of torsion and nonmetricity are affected. A similar logic applies to the axial-torsion as well, though in this case we are dealing with a pseudo-vector in the physical case ($d=4$).
The complete decomposition of the torsion involves a vector, an axial-tensor, which becomes an axial vector in the physical case, and a tensor
\begin{equation}\label{torsion-decomposition}
\begin{split}
 T^\nu{}_{\rho\mu} & =  \frac{1}{d-1} \bigl( \delta^\nu{}_\mu \tau_\rho
 - \delta^\nu{}_\rho \tau_\mu \bigr)
 \\&
 - \frac{1}{3!(d-3)!} \varepsilon^{\sigma_1 \cdots \sigma_{d-3}\nu}{}_{\rho\mu} \theta_{\sigma_1 \cdots \sigma_{d-3}} + \kappa^\nu{}_{\rho\mu} \, .
\end{split}
\end{equation}
We define the first two as
the torsion vector $\tau_\rho = T^\nu{}_{\rho\nu}$ and the axial-torsion
$\theta_{\sigma_1 \cdots \sigma_{d-3}} = T^\nu{}_{\rho\mu} \varepsilon_{\sigma_1 \cdots \sigma_{d-3}\nu}{}^{\rho\mu}$.
It is easy to deduce the transformation properties of all the (pseudo-)vectors of a general MAG
\begin{equation}\label{proj-MAGvectors-general}
\begin{split}
 \delta \tau_\rho & =   (d-1) (p_\rho - q_\rho) \, ,\\
 \delta \theta_{\sigma_1 \cdots \sigma_{d-3}} &=
 c_d  s_{\sigma_1 \cdots \sigma_{d-3}} \,, \\
 \delta Q_\mu{}^\nu{}_\nu &=   2 d \, p_\mu + 2 (q + r)_\mu \, ,\\
 \delta Q_\mu{}^\mu{}_\rho &=   2 \, p_\rho + (d+1) (q + r)_\rho \, ,
\end{split}
\end{equation}
having defined $ c_d= 12(d-3)! $ (the sign of the transformation of $\theta$ depends on the signature, which is chosen to be Lorentzian here).
In the physical case of four dimension, the formulas specialize to
\begin{equation}\label{proj-MAGvectors}
\begin{split}
 \delta \tau_\rho  &=  3 (p_\rho - q_\rho) \, ,\\
 \delta \theta_\sigma  &=  12 s_\sigma \, ,\\
 \delta Q_\mu{}^\nu{}_\nu &=  8 p_\mu + 2 q_\mu + 2 r_\mu \, ,\\
 \delta Q_\mu{}^\mu{}_\rho &=  2 p_\rho + 5 q_\rho + 5 r_\rho \, .
\end{split}
\end{equation}

\section{Weyl gauging and invariant contortion}\label{sect:weyl}

The spin-connection can be seen as a gauge connection of the Lorentz group $SO(1,d-1)$
which is a subgroup of the general $GL(d)$ transformations \cite{Percacci:1990wy}.
Another natural subgroup of $GL(d)$ includes the generator of the dilatations, as we briefly clarify in Appendix~\ref{sect:gld}. Local scale transformations are Weyl transformations of the metric, which naturally transforms with weight $w(g_{\mu\nu})=2$
as $g_{\mu\nu} \to g'_{\mu\nu}=e^{2 \sigma} g_{\mu\nu}$ with $\sigma$ a local function over spacetime.
We can construct a connection which is both generally covariant and Weyl gauge covariant as follows
\begin{equation}\label{eq:weyl-gauged-connection}
 \hat{\nabla}_\mu v^\nu = \mathring{\nabla}_\mu v^\nu + L^\nu{}_{\rho\mu} v^\rho + w_v S_\mu v^\nu\,,
\end{equation}
where $w_v$ is the Weyl weight of the vector $v^\mu$, $S_\mu$ is the gauge potential
of the local scale transformations (Weyl gauge potential), and $L^\nu{}_{\rho\mu}= \delta^\nu_\rho S_\mu+\delta^\nu_\mu S_\rho-g_{\rho\mu} S^\nu$ \cite{Charap:1973fi,Iorio:1996ad,Sauro:2022chz}.
The covariant derivative $\hat{\nabla}$ is such that, under the local transformations
\begin{equation}
\begin{split}
 g_{\mu\nu} & \to g'_{\mu\nu} =e^{2 \sigma} g_{\mu\nu}\,,
 \\
 v^\mu & \to v^{\prime\mu}=e^{w_v \sigma} v^\mu\,,
 \\
 S_\mu & \to S'_\mu=S_\mu -\partial_\mu \sigma\,,
\end{split}
\end{equation}
it transforms covariantly as $v^\mu$ itself
\begin{equation}
 \hat{\nabla}_\mu v^\nu \to \hat{\nabla}'_\mu v^{\prime\nu}=e^{w_v \sigma} \hat{\nabla}_\mu v^\nu\,.
\end{equation}
One finds that $L^\nu{}_{\rho\mu}$ is a necessary additional contribution that balances the nontrivial Weyl transformation of $\mathring{\nabla}$ \cite{Iorio:1996ad}.
The additional contribution can be seen as extending the original Levi-Civita connection in the realm of MAG (e.g., compare \eqref{eq:weyl-gauged-connection} with $w_v=0$
and \eqref{eq:Gamma}). In practice, the additional contribution can be seen as motivated
by the fact that the generator of the dilatations does not commute with the other generators of the spacetime symmetries (see Eq.~\eqref{eq:commutators-dilatations} in Appendix~\ref{sect:gld} for the explicit relation).

The Weyl weight is essentially the negative of the scaling dimension of the given field. Eq.~\eqref{eq:weyl-gauged-connection} generalizes as
\begin{equation}\label{eq:weyl-gauged-connection-general}
 \hat{\nabla}_\mu {\cal V}^A = \mathring{\nabla}_\mu {\cal V}^A + (L_{\mu})^A{}_B {\cal V}^B + w_{{\cal V}} S_\mu {\cal V}^A\,,
\end{equation}
where ${\cal V}^A$ is an arbitrary tensor, $w_{{\cal V}}$ its weight, and $(L_{\mu})^A{}_B$ the contraction of the components of $L$ with the appropriate generators.
To see that the action of the connection
has the expected behavior on the Weyl weight, it is convenient to recall that the vielbein $e^a{}_\mu$ must have weight $w(e^a{}_\mu)=1$ to be consistent with the weight of the metric and with the fact that $\eta_{ab}$ must not transform; then, if we define $\hat{\nabla}_a v^\nu = E^\mu{}_a \hat{\nabla}_\mu v^\nu$, the transformation becomes
\begin{equation}
 \hat{\nabla}_a {\cal V}^A \to \hat{\nabla}'_a {\cal V}^{\prime A}=e^{(w_v-1) \sigma} \hat{\nabla}_a {\cal V}^A\,,
\end{equation}
where we used the fact that the inverse vielbein has weight $w(E^\mu{}_a)=-1$.
It is straightforward to see that, likewise $\mathring{\nabla}$, $\hat{\nabla}$ is compatible with the metric
\begin{equation}
 \hat{\nabla}_\mu g_{\nu\rho} =
 -L^\theta{}_{\nu\mu} g_{\theta\rho}-L^\theta{}_{\rho\mu} g_{\nu\theta} + 2 S_\mu g_{\nu\rho} =0\,,
\end{equation}
where we used the fact that $\mathring{\nabla}_\mu g_{\nu\rho}=0$ and the explicit form of $L^\nu{}_{\rho\mu}$ as a function of $S_\mu$. Noticing that $\hat{\nabla}$ is symmetric, it should be clear that $\hat{\nabla}$ could emerge dynamically as a general solution to the Palatini formulation of gravity with an Einstein-Hilbert action \cite{Wheeler:2022ggm,Sauro:2022chz}.
An amusing feature of $\hat{\nabla}$, related to its metric-compatibility, is that covariant integration by parts happens flawlessly as long as the argument of integration
is a scalar with weight $d$ \cite{Sauro:2022chz}. In the mathematical literature, a manifold equipped with the pair $(g_{\mu\nu},\hat{\nabla})$ is often referred as a Weyl or \emph{Weylian geometry},
as opposed to the \emph{Riemannian geometry} given by the pair $(g_{\mu\nu},\mathring{\nabla})$ \cite{ehlers1973geometry,Scholz:2011za}.

From the point of view of MAG, the contribution $L^\nu{}_{\rho\mu}$ is a special kind of distortion term, whereas the connection $\hat{\nabla}$ gives rise to vector torsion only when acting on Weyl-charged fields.
In Ref.~\cite{Sauro:2022chz} we have shown how to add a torsional part to the connection
\eqref{eq:weyl-gauged-connection-general} starting from a general MAG connection
and discussing the invariant decomposition of the action of dilatations.
The final result of the analysis of \cite{Sauro:2022chz} has been the construction of a \emph{Riemann-Cartan-Weyl connection}, defined as
\begin{equation} \label{eq:weyl-gauged-connection-general-with-torsion}
\begin{split}
 \tilde{\nabla}_\mu {\cal V}^A
 & = \hat{\nabla}_\mu {\cal V}^A
 + (\hat{K}_{\mu})^A{}_B {\cal V}^B \\
 & =
 \mathring{\nabla}_\mu {\cal V}^A + (\hat{K}_{\mu})^A{}_B {\cal V}^B + (L_{\mu})^A{}_B {\cal V}^B
 \\& + w_{{\cal V}} S_\nu {\cal V}^A\,,
\end{split}
\end{equation}
where $\hat{K}_\mu$ is a Weyl \emph{invariant} contortion with components $\hat{K}^\nu{}_{\rho\mu}$. The connection \eqref{eq:weyl-gauged-connection-general-with-torsion}
can be interpreted from the point of view of a general MAG as in \eqref{eq:Gamma}
as having special contortion $\hat{K}$ and disformation $L$, besides the additional Abelian weight $w_{\cal V}$.

The connection $\tilde{\nabla}$ differs from a completely general metric-affine connection in that the distortion tensor is a pure trace, which is constructed with a single $1$-form, thus its two independent traces are linearly dependent. Moreover, the requirement of Weyl invariance automatically makes the number of possible interaction terms of the most general action functional finite. This is not true in a general metric-affine theory, for in such a case a Lagrange density cubic in the curvature scalar would be allowed. In Appendix~\ref{sect:coset} we give a coset-based
derivation of this covariant derivative, as well as all other covariant derivatives that have appeared in the previous sections \cite{Karananas:2015eha}.

Given that, as we have seen in the previous section, projective transformations do not
alter the tensorial contributions to contortion and disformation due to their vectorial nature, it makes sense to investigate the effect of the general projective transformations and decompose them according to the Weylian structures that we have just discussed.

\section{Projective transformations in Riemann-Cartan-Weyl gravity}\label{sect:weyl-projective}

%
To begin with, we need to identify the affine potentials that must transform under
the action of the projective transformations.
We use the tetrad postulate, $\tilde{\nabla}e=0$, so that the $GL(d)$ connection
is expressed in terms of the spin-connection and the Weyl potential
\begin{equation}\label{aff-conn-CW}
\Gamma^\nu{}_{\rho\mu} = E^\nu{}_a \left( \partial_\mu e^a{}_\rho + \omega^a{}_{b\mu} e^b{}_\rho + S_\mu e^a{}_\rho \right) \, .
\end{equation}
There are in general two possible affine transformations that preserve
the antisymmetric property of $\omega^a{}_{b\mu}$
\begin{equation}\label{proj-mag-general-CW-ansatz}
\begin{split}
 S_\mu &\to  S_\mu + \xi_\mu \,, \\
 \omega^a{}_{b\mu}
 & \to  \omega^a{}_{b\mu} + \left( e^a{}_\mu E^\nu{}_b - e_{b\mu} E^{\nu a}  \right) v_\nu
 \\
 & +  E^{\nu_1}{}_{f_1} \cdots E^{\nu_{d-3}}{}_{f_{d-3}} e^c{}_\mu  \varepsilon^{f_1 \cdots f_{d-3} a}{}_{bc}  w_{\nu_1 \cdots }\,,
\end{split}
\end{equation}
and that are parametrized by two independent vectors $\xi_\mu$ and $v_\nu$, and a totally antisymmetric tensor $w_{\nu_1 \cdots \nu_{d-3}}$ with $d-3$ indices.
Likewise the previous section, we have that, in $d=4$, $w_\mu$ becomes an axial vector, and, in $d=3$, it becomes a pseudoscalar.\footnote{%
An important comment is in order. Given that $\omega^a{}_{b\mu}$ has Weyl weight zero, we deduce that $\xi_\mu$ and $v_\mu$ have weight zero. Instead, $w_{\nu_1 \cdots \nu_{d-3}}$ has Weyl weight $d-4$ because $\varepsilon$ with anholonomic indices has weight zero. This implies that there is a ``semidirect'' action between
the Weyl and the axial projective transformations.
}

Combining the transformations together, we see that
\begin{equation}\label{proj-mag-general-CW}
\begin{split}
 \Gamma^\nu{}_{\rho\mu} & \to \Gamma^\nu{}_{\rho\mu} + \delta^\nu{}_\rho \xi_\mu + \left( \delta^\nu{}_\mu v_\rho - g_{\mu\rho} v^\nu \right)
 \\&
 +  \varepsilon^{\sigma_1 \cdots \sigma_{d-3}\nu}{}_{\rho\mu} w_{\sigma_1 \cdots \sigma_{d-3}}\,,
\end{split}
\end{equation}
and, comparing with \eqref{proj-mag-general}, we have the obvious identifications $p_\mu=\xi_\mu$, $q_\mu=-r_\mu=v_\mu$ and $s_{\sigma_1 \cdots \sigma_{d-3}}=w_{\sigma_1 \cdots \sigma_{d-3}}$. While in a general MAG we have four independent vector parameters for the projective transformations, in the
Riemann-Cartan-Weyl formulation there are only three, because the Riemann-Cartan-Weyl nonmetricity is of the Weyl form, implying that the traces $Q_\mu{}^\nu{}_\nu$ and $Q_\nu{}^\nu{}_\mu$ are linearly dependent.

In the following we will be interested in the two truly vectorial projective transformations of \eqref{proj-mag-general-CW}, so we drop the transformation associated
to the tensor $w_{\sigma_1 \cdots \sigma_{d-3}}$. We define the \emph{first Cartan-Weyl projective transformation} as the one parametrized by $\xi_\mu$ in \eqref{proj-mag-general-CW-ansatz}
\begin{equation}\label{eq:projective-cw-1}
\begin{split}
 \delta^{p_1}_\xi S_\mu = \xi_\mu\,, \quad \delta^{p_1}_\xi\omega^a{}_{b\mu}=0
 \,, \quad \delta^{p_1}_\xi g_{\mu\nu}=0\,,
\end{split}
\end{equation}
and the \emph{second Cartan-Weyl projective transformation} as the one parametrized by $v_\mu$ in \eqref{proj-mag-general-CW-ansatz}
\begin{equation}\label{eq:projective-cw-2}
\begin{split}
 &\delta^{p_2}_v S_\mu = 0\,, \quad \delta^{p_2}_v\omega^a{}_{b\mu}=\left( e^a{}_\mu E^\nu{}_b - e_{b\mu} E^{\nu a}  \right) v_\nu
 \,,
 \\&
 \delta^{p_2}_v g_{\mu\nu}=0\,.
\end{split}
\end{equation}
These transformations can be though of as finite or infinitesimal, depending on the generators being finite or infinitesimal themselves.

Suppose now that we have an action that depends on the geometric structures of the Riemann-Cartan-Weyl geometry. For simplicity we assume that it does not depend on
the axial torsion tensor, which allows us to avoid some unnecessary complications
since we are working in arbitrary $d$.
From the point of view of \eqref{proj-mag-general-CW-ansatz} it \emph{seems} that, using the first transformation, we can change $S_\mu$ at will with a specific choice of the parameter ($\xi_\mu=-S_\mu$), thus eliminating in each local chart the Weyl gauge potential $S_\mu$. As a consequence, one would be tempted to assume that requiring invariance under the first transformation decouples $S_\mu$ from the theory, making
the resulting model ``conformal'' in the sense that it would be invariant under Weyl transformations (not gauged). This, however, is not completely true, because the spin-connection depends on both torsion and potential $S_\mu$. The torsion vector \emph{does}
transform under the first transformation \eqref{proj-mag-general-CW-ansatz}, implying that the dependence on $S_\mu$ is inherited by the torsion vector itself.

To see how this happens in practice, take the definition of the torsion vector, $\tau_\rho = T^\nu{}_{\rho\nu}$. Using \eqref{eq:projective-cw-1} and \eqref{eq:projective-cw-2}, we see that
\begin{equation}
 \delta^{p_1}_\xi \tau_\mu = (d-1) \xi_\mu \,,\qquad
 \delta^{p_2}_v \tau_\mu = -(d-1) v_\mu \, .
\end{equation}
We have that the combination $S_\mu-(d-1)\tau_\mu$ of Weyl gauge potential and torsion vector is invariant under the first Cartan-Weyl projective transformation
\begin{equation}
 \delta^{p_1}_\xi \bigl( S_\mu-\frac{1}{d-1}\tau_\mu \bigr) =0\,,
\end{equation}
while the torsion vector is left invariant by a special linear combination of the
two Cartan-Weyl projective transformations
\begin{equation}\label{eq:projective-cw-extra}
 \bigl(\delta^{p_1}_\xi+\delta^{p_2}_\xi \bigr) \tau_\mu =0\,, \qquad
 \bigl(\delta^{p_1}_\xi+\delta^{p_2}_\xi \bigr) S_\mu = \xi_\mu\,.
\end{equation}
For convenience we denote the linear combination of the two transformations as
$\delta^{p_0}_\xi=\delta^{p_1}_\xi+\delta^{p_2}_\xi$.

Now consider the action $S[\psi,g_{\mu\nu},\tau_\mu,S_\mu]$ of a theory describing some field $\psi$ coupled to the Riemann-Cartan-Weyl geometry. We omitted further arguments, which could include the remaining components of the torsion field.
The invariance under the projective transformations have several implications for the action.
For example, invariance of the action under the first transformation, $\delta^{p_1}_\xi S[\psi,g_{\mu\nu},\tau_\mu,S_\mu]=0$, implies that the action is actually
a functional of the combination $S_\mu-\frac{1}{d-1}\tau_\mu$.
Thus, if we choose the ``gauge'' for which $\xi_\mu=-S_\mu$, then the resulting torsion vector is modified by $S_\mu$. In order to maintain the actual form of the torsion vector, we
require invariance under the linear combination, $\delta^{p_0}_\xi S[\psi,g_{\mu\nu},\tau_\mu,S_\mu]=0$, so we can make the ``gauge'' choice $\xi_\mu=-S_\mu$
to write the action as $S[\psi,g_{\mu\nu},\tau_\mu,0]$ where now $\frac{1}{d-1}\tau_\mu$ plays the role of the Weyl gauge potential. We refer to the resulting theory as ``torsion gauged'', as the torsion plays the role of the Weyl-gauge potential.
Finally, requiring invariance under $\delta^{p_2}_\xi$
separately, we can further ``gauge'' away the torsion vector too, and the theory is completely independent of both $S_\mu$ and $\tau_\mu$, implying that it is a Weyl-invariant theory in the normal nongauged sense (conformal).

A summary of the effects of the projective transformations and their outcomes on
general Riemann-Cartan-Weyl theories is given by the following sequence
\begin{equation*}
\begin{split}
 &{\rm Riemann-Cartan-Weyl}
 \\&
 \quad\overset{\delta^{p_0}}{\longrightarrow}
 \quad
 {\rm Torsion-gauged}
 \\&
 \quad\overset{\delta^{p_2}}{\longrightarrow}
 \quad
 {\rm Conformal} \, ,
 \end{split}
\end{equation*}
where a torsion-gauged theory is essentially the model discussed in \cite{Karananas:2015eha,Karananas:2021gco},
and a conformal theory is one that is invariant under Weyl transformations without the aid of a gauge potential.

\section{Example: scalar theory in any $d$}\label{sect:scalar-d-general}

As a first example, we want to discuss a scalar theory coupled to a Riemann-Cartan-Weyl geometry in arbitrary $d$. For simplicity we leave aside the axial torsion, which we return to when specializing to the $d=4$ case, so we assume $\theta_\mu=0$ temporarily.

It is convenient to express everything in terms of
anholonomic indices, i.e., exploiting the local Lorentz formulation.
Using a canonically normalized scalar field $\varphi$ with Weyl-weight $w_\varphi=\frac{2-d}{2}$, the covariant derivative $\tilde{\nabla}_a = E^\mu{}_a \tilde{\nabla}_\mu$, the curvature of $\tilde{\nabla}$,
and the torsion vector $\tau^a= e^\mu{}_a \tau^\mu$,
we can construct the following Lagrangian terms on the basis of dimensional analysis
\begin{equation}\label{eq:scalar-terms-general-d}
 \tilde{\nabla}_a \varphi \tilde{\nabla}^a \varphi\,,
 \quad
 \varphi^2 \tau^a\tau_a\,,
 \quad
 \varphi^2 \tilde{R}\,,
 \quad
 \tau^a \tilde{\nabla}_a \varphi\,,
\end{equation}
where $\tilde{R}= \eta^{bc}E^\nu{}_cE^\mu{}_a\tilde{R}^a{}_{b\mu\nu}$.
The appropriate invariant integration element is
$\underline{e} = \det e^a{}_\mu$.
The linear combination of the terms of \eqref{eq:scalar-terms-general-d}
that is invariant under the first and second projective transformations,
given in Eqs.~\eqref{eq:projective-cw-1} and \eqref{eq:projective-cw-2} respectively, is
\begin{equation}\label{proj-inv-action1}
\begin{split}
 S &= -\frac{1}{2} \int \underline{e} \, \Bigl\{ (\tilde{\nabla}_a \varphi)^2 + \frac{(d-2)^2}{4 (d-1)^2} \tau^2 \varphi^2
 \\&
 + \frac{d-2}{2(d-1)}  \tau^a \tilde{\nabla}_a \varphi^2 + \frac{d-2}{4 (d-1)} \varphi^2 \tilde{R} \Bigr\}\,,
\end{split}
\end{equation}
which we normalized with an overall canonical factor $\frac{1}{2}$.
Notice that the connection $\tilde{\nabla}$ contains the torsion,
so it cannot be integrated by parts straightforwardly, even though it is compatible with the metric. Manipulations of this sort are simplified by adopting the covariant derivative $\hat{\nabla}$, defined in \eqref{eq:weyl-gauged-connection-general},
which \emph{can} be integrated by parts \emph{if and only if} the integrand has dimension $d$.\footnote{%
In doubt, the safest procedure is to express everything in terms of $\mathring{\nabla}$, which can always be integrated by parts no matter what.
}
Some important transformations used for the derivation of \eqref{proj-inv-action1} are summarized in Appendix~\ref{sect:variations}.

A simple, but tedious, self-consistency check is to observe that \eqref{proj-inv-action1}
depends on $S_\mu$ and $\tau_\mu$ only through the combination
$S_\mu-\frac{1}{d-1}\tau_\mu$, so, using \eqref{eq:projective-cw-extra} for $\xi_\mu=-S_\mu$, we can rewrite it as a scalar Weyl-gauged action in which the role of the Weyl gauge potential is played by the torsion vector (as discussed at the end of the previous section).
It is also particularly simple to see that, eliminating also
the torsion vector, the action \eqref{proj-inv-action1} \emph{almost}
reduces to the standard Weyl-invariant (conformal) action with kernel the Weyl-covariant Yamabe operator $-\mathring{\nabla}^2+\frac{d-2}{4(d-1)}\mathring{R}$. The difference with the standard Yamabe action, besides a boundary term, is that $\tilde{R}$
contains contributions from the truly tensorial and Weyl-invariant part of the torsion.

We are not done yet. As previously stated, the truly tensorial part of the torsion $\kappa^a{}_{bc}$, seen in \eqref{torsion-decomposition}, is Weyl-invariant.
Therefore, to the list of terms in \eqref{eq:scalar-terms-general-d}
we can freely add the interaction $\varphi^2 \kappa^a{}_{bc} \kappa_a{}^{bc}$,
which is the unique scalar combination compatible with the mixed symmetries of $\kappa^a{}_{bc}$ and with the correct Weyl weight. The interaction strength $\zeta$ of this term is thus not constrained by the projective symmetries and the action \eqref{proj-inv-action1} is complemented by
\begin{equation}\label{proj-inv-action1.2}
\begin{split}
 \Delta S &= -\frac{1}{2} \int \underline{e} \, \zeta \varphi^2 \kappa^a{}_{bc} \kappa_a{}^{bc} \,.
\end{split}
\end{equation}

\section{Example: scalar theory in $d=4$ and the axial torsion}\label{sect:scalar-4d}

Now we specialize the result of the previous section to $d=4$, but we also add the axial torsion, which is a vector in the physical dimension.
As for the transformation of the axial torsion, we first specialize \eqref{proj-mag-general-CW-ansatz} and \eqref{proj-mag-general-CW} to $d=4$. The \emph{axial Cartan-Weyl projective transformation} is defined
\begin{equation}\label{eq:projective-cw-axial}
\begin{split}
 &\delta^{p_3}_w S_\mu = 0\,, \quad \delta^{p_3}_w\omega^a{}_{b\mu}=E^{\nu}{}_{f}  e^c{}_\mu  \varepsilon^{f a}{}_{bc}  w_{\nu}
 \,,
 \\& \delta^{p_3}_w g_{\mu\nu}=0\,,
\end{split}
\end{equation}
and the generator $w_\mu$ has the same weight as the generators of the other two projective transformations \eqref{eq:projective-cw-1} and \eqref{eq:projective-cw-2}
thanks to the fact that we are in four dimensions.
Using the definitions, we see that the vector torsion is unchanged, while the axial torsion is transformed
\begin{equation}
 \delta^{p_3}_v \tau_\mu = 0 \,,\qquad
 \delta^{p_3}_v \theta_\mu = 12 w_\mu \, ,
\end{equation}
as expected. We also have that the axial torsion does not change under the other transformations $\delta^{p_3}_v \tau_\mu=0$.

The available monomials given in \eqref{eq:scalar-terms-general-d} remain, but must be complemented with the term $\varphi^2 \theta^2=\varphi^2 \theta_a \theta^a$, instead $\theta^a \tilde{\nabla}_a \varphi^2$ is excluded because of parity. The requirement that the action is invariant under the first two projective transformations, \eqref{eq:projective-cw-1} and \eqref{eq:projective-cw-2}, as well as \eqref{eq:projective-cw-axial}, results in
\begin{equation}\label{proj-inv-action2}
\begin{split}
 S &= -\frac{1}{2} \int \underline{e} \, \Bigl\{ (\tilde{\nabla}_a \varphi)^2 + \frac{1}{9} \tau^2 \varphi^2 + \frac{1}{3}  \tau^a \tilde{\nabla}_a \varphi^2
 \\&
 + \frac{1}{144} \theta^2 \varphi^2
 + \frac{1}{6} \varphi^2 \tilde{R}
 + \zeta \varphi^2 (\kappa^a{}_{bc})^2
 +\frac{\lambda}{12} \varphi^4
 \Bigr\}\,,
\end{split}
\end{equation}
to which we added the scalar self-interaction that is invariant in $d=4$.
If we express everything in terms of the Levi-Civita connection and use the three projective symmetries to explicitly eliminate $S_\mu$, $\tau_\mu$ and $\theta_\mu$, we get
\begin{equation}\label{proj-inv-action2.2}
\begin{split}
 S = &-\frac{1}{2} \int \underline{e} \, \Bigl\{ (\mathring{\nabla}_a \varphi)^2
 + \frac{1}{6} \varphi^2 \mathring{R}
 \\&
 + \Bigl(\frac{1}{2}+\zeta\Bigr) \varphi^2 \kappa^a{}_{bc} \kappa_a{}^{bc}
 +\frac{\lambda}{12} \varphi^4
 \Bigr\}\,.
\end{split}
\end{equation}
Thus, the action functional for a scalar field projectively coupled to Riemann-Cartan-Weyl geometry is described in terms of two free parameters only.
It would be interesting to explore the dependence on $\kappa^a{}_{bc}$ of the conformal anomaly.

\section{Comments on local Weyl invariance}\label{sect:comments}

\subsection{The elimination of the Weyl-gauge potential and torsion-gauging}

Here we would like to shine some light on the conformal behavior of the projective-invariant actions \eqref{proj-inv-action1} and \eqref{proj-inv-action2}.
Reading the Appendices~\ref{sect:gld} and~\ref{sect:coset} could be useful either at this point or after this section, but not mandatory.

The first, rather mundane, comment is that we might have expected a conformal behavior from the very beginning, since we have been preserving the light-cone projective-structure, which is tightly connected with conformal invariance for obvious reasons (conformal transformations preserve the light-cone).
A sufficient condition for the realization of conformal symmetry in an action functional is invariance under local Weyl transformations without the need for the gauge potential $S_\mu$. In this sense, local Weyl invariance can be thought of as Weyl gauge invariance with gauge potential $S_\mu=0$ (i.e., it is decoupled and therefore can be set to zero consistently without undermining Weyl invariance). This last condition can be recovered by acting with the first projective transformation \eqref{eq:projective-cw-1} with the parameter $\xi_\mu$ chosen as $ \xi_\mu=-S_\mu$.

Let us consider the torsion $2$-form geometrically, defined as the exterior covariant derivative of the vielbein forms $e^a = e^a{}_\mu dx^\mu$
\begin{align}
\mathcal{T}^a &= \, d e^a + \omega^a{}_b \wedge e^b + S \wedge e^a \, .
\end{align}
As we have discussed recently in Ref.~\cite{Sauro:2022chz}, there are two ``natural'' ways to split the spin-connection into torsion-free and torsion-full terms
\begin{equation}
\omega^a{}_b = \mathring{\omega}^a{}_b + \check{\Omega}^a{}_b = \hat{\omega}^a{}_b + \hat{\Omega}^a{}_b \, ,
\end{equation}
where $\mathring{\omega}^a{}_{b\mu}= E^\nu{}_b \bigl( \mathring{\Gamma}^\rho{}_{\nu\mu} e^a{}_\rho - \partial_\mu e^a{}_\nu \bigr)$ is the torsion-free spin-connection associated to the Levi-Civita connection also seen in the previous sections.
The relation between the two splits is given by
\begin{equation}
\hat{\omega}^a{}_b = \mathring{\omega}^a{}_b + e^a S_b - e_b S^a \, .
\end{equation}
The connection $\hat{\omega}$ can be defined geometrically by considering a Weyl-covariant exterior derivative $\overset{w}{D} = d + w S$. It acts on the co-frame in the defining equation
\begin{equation}
\hat{\omega}^{ab}
 \equiv   \frac{1}{2} E^b \lfloor \, \overset{w}{D} e^a - \frac{1}{2} E^a \lfloor \, \overset{w}{D} e^b + \frac{1}{2} e_c E^a \lfloor E^b \lfloor \, \overset{w}{D} e^c \, ,
\end{equation}
where we adopted the floor operator which contracts contravariant indices with covariant ones from the left.
It is important to stress that the full spin-connection of this approach is exactly Weyl invariant from the onset, as discussed in Ref.~\cite{Sauro:2022chz}. The torsion $2$-form is Weyl covariant in either way and its explicit expressions in the two approaches are
\begin{equation}
 \mathcal{T}^a  = \check{\Omega}^a{}_b \wedge e^b + S \wedge e^a  = \hat{\Omega}^a{}_b \wedge e^b \, .
\end{equation}

Let us now consider the alternate situation in which local Weyl symmetry is realized without the need of Weyl potential. We focus on the former decomposition of the spin-connection and of the torsion $2$-form accordingly. We have that $\mathcal{T}^a$ has a nonhomogeneous Weyl transformation law in the nongauged approach, which, however, affects only its vector irreducible component. Let us define a torsion-vector associated to $\check{\Omega}^a{}_b$
\begin{equation}
\check{\tau}_a = E^\mu{}_b \, \check{\Omega}^b{}_{a \mu} \, .
\end{equation}
Its infinitesimal Weyl transformation with parameter $\sigma$, denoted $\delta^W_{\sigma}$, is nonhomogeneous
\begin{equation}\label{Weyl-var-tau-check}
\delta^W_{\sigma} \check{\tau}_a = - \sigma \check{\tau}_a - (d-1) \partial_a \sigma \, .
\end{equation}
Obviously, the gradient of a scalar field $\varphi$ transforms as
\begin{equation}\label{Weyl-var-grad-phi}
\delta^W_{\sigma} \partial_a \varphi = - \frac{d}{2} \sigma \partial_a \varphi - \frac{d-2}{2} \varphi \partial_a \sigma \, ,
\end{equation}
which also includes a nonhomogeneous part.
After using the first projective transformation to ``eliminate'' the Weyl potential $S_\mu$, the part of the Lagrangian which depends on the the derivatives of the scalar field and on $\check{\tau}_a$ is
\begin{equation}\label{phi-tau-check-Lagrangian}
\begin{split}
{\cal L} \supset & -\frac{1}{2}\Bigl\{
(\partial_a \varphi)^2 + \frac{(d-2)^2}{4(d-1)^2} \check{\tau}^2 \varphi^2
\\&
+ \frac{d-2}{2(d-1)} \check{\tau}^a \partial_a \varphi^2
\Bigr\}\, .
\end{split}
\end{equation}
We could define the following ``check'' derivative of the scalar field
\begin{equation}
 \check{D}_a \varphi = \partial_a \varphi + \frac{d-2}{2(d-1)} \check{\tau}_a \varphi \, .
\end{equation}
Using Eqs.~\eqref{Weyl-var-tau-check} and \eqref{Weyl-var-grad-phi}, we find that the Weyl-variation of this derivative is homogeneous
\begin{equation}
 \delta^W_\sigma \check{D}_a \varphi = - \frac{d}{2} \sigma \check{D}_a \varphi \, .
\end{equation}
As expected, the torsion-vector plays the role of a Weyl potential and allows us to define a Weyl covariant derivative. This approach towards Weyl-invariance is sometimes adopted in the metric-affine literature, most notably in \cite{Karananas:2021gco}.
It is straightforward to see that the Lagrangian term \eqref{phi-tau-check-Lagrangian}
can be rewritten as the square of the new Weyl-covariant derivative of the scalar field
\begin{equation}
 {\cal L} \supset -\frac{1}{2}(\check{D}_a \varphi)^2 \, .
\end{equation}

Going back to the full action \eqref{proj-inv-action1}, let us focus on the interaction between the scalar and the curvature. As remarked above, the full spin-connection is exactly Weyl-invariant, so $\tilde{R}$ is Weyl-covariant with weight $-2$. This is because the affine transformation of the Levi-Civita scalar curvature $\mathring{R}$ is exactly balanced by the terms which involve $\check{\Omega}^a{}_b$. Thus, the Lagrangian
\begin{align}
{\cal L} \supset -\frac{1}{2}\Bigl\{  (\check{D}_a \varphi)^2 + \frac{d-2}{4(d-1)} R \, \varphi^2 \Bigr\} \,
\end{align}
gives a locally Weyl invariant action. This property still holds true if we impose, in general only at the kinematical level, the torsion-free condition; in which case the Lagrangian eventually reads
\begin{equation}
{\cal L} \supset -\frac{1}{2}\Bigl\{ (\partial_a \varphi)^2 + \frac{d-2}{4(d-1)} \mathring{R} \varphi^2 \Bigr\} \, .
\end{equation}
This Lagrangian is precisely that of a scalar field conformally-coupled to the Ricci scalar.

\subsection{Formally reducing the field variables: bosonic case}

As we have done multiple times above, we can exploit the projective invariances of the complete action functional to rearrange the dependence on the field variables. Here we would like to outline this procedure for gravity coupled to bosonic fields.
Generically, the action functional can have the following dependencies
\begin{equation}
S= S[\phi, g_{\mu\nu}, \Gamma^\rho{}_{\nu\mu}, S_\mu] \, ,
\end{equation}
where $\Gamma^\rho{}_{\nu\mu}$ is the full holonomic affine-connection and $\phi$ is some unspecified bosonic field. The action might depend explicitly on the curvature as well as the torsion of $\Gamma_\mu$, and this would not alter our considerations.

We choose the Weyl-affine splitting of the spin-connection
\begin{equation}
\omega^a{}_{b\mu} = \mathring{\omega}^a{}_{b\mu} + \check{\Omega}^a{}_{b\mu} \, ,
\end{equation}
which induces the analogous Weyl-affine splitting of the holonomic connection \cite{Sauro:2022chz}
\begin{equation}
\Gamma^\rho{}_{\nu\mu} = \mathring{\Gamma}^\rho{}_{\nu\mu} + \check{K}^\rho{}_{\nu\mu} + \delta^\rho{}_\nu S_\mu \, .
\end{equation}
Exploiting this decomposition the dependencies of the bosonic action functional become
\begin{equation}
S = S[\phi, g_{\mu\nu}, \check{K}^\rho{}_{\nu\mu}, S_\mu] \, .
\end{equation}

Let us analyze the consequences of imposing the invariance of the action under the first two projective transformations \eqref{eq:projective-cw-1} and \eqref{eq:projective-cw-2}. Since the projective parameters $\xi_\mu$ and $v_\mu$ are continuous, we can derive two associated N\"other identities. To this end, we first define the following tensors
\begin{equation}
\begin{split}
 & T^{\mu\nu} =  \frac{2}{\sqrt{-g}} \frac{\delta S}{\delta g_{\mu\nu}} \, ,
 \quad
 \Sigma_\rho{}^{\nu\mu} = \frac{1}{\sqrt{-g}} \frac{\delta S}{\delta \check{K}^\rho{}_{\nu\mu}}\,,
 \\&
 \Delta^\mu = \frac{1}{\sqrt{-g}} \frac{\delta S}{\delta S_\mu} \, ,
\end{split}
\end{equation}
which we refer to as \emph{energy-momentum}, \emph{spin-current} and \emph{dilation-current} tensors, respectively.
Using the gravitational field variables $g_{\mu\nu}$, $\check{K}^\rho{}_{\nu\mu}$ and $S_\mu$, the first projective transformation \eqref{eq:projective-cw-1} acts non-trivially only on the Weyl potential. Thus, the variation of the action reads
\begin{equation}
\begin{split}
 \delta^{p_1}_\xi S[\phi, g_{\mu\nu}, \check{K}^\rho{}_{\nu\mu}, S_\mu]
 &=  \int \sqrt{-g} \Delta^\mu \delta^{p_1}_\xi S_\mu
 \\
 &= \int \sqrt{-g} \Delta^\mu \xi_\mu \, .
\end{split}
\end{equation}
Given the arbitrariness of the parameter $\xi_\mu$, the projective invariance of the first type yields $\Delta^\mu = 0$
as the necessary condition for the action functional to be invariant under the first projective transformation.

Now we require $\delta^{p_2}_v S = 0$. The second projective transformation acts on the vector part of $\check{K}^\rho{}_{\nu\mu}$ only, i.e., on $\check{\tau}_\mu = - \check{K}^\nu{}_{\mu\nu}$. In order to single out the contribution, we decompose the Weyl-affine contortion tensor as
\begin{equation}
\check{K}^\rho{}_{\nu\mu} = - \frac{1}{d-1} \left( \delta^\rho{}_\mu \check{\tau}_\nu - g_{\mu\nu} \check{\tau}^\rho \right) + \kappa^\rho{}_{\nu\mu} \, ,
\end{equation}
where all the information about the axial-vector and tensor irreducible components of torsion are hidden in $\kappa^\rho{}_{\nu\mu}$ on the right-hand-side (we include both contributions in $\kappa$ for notational simplicity, so $\kappa$ here contains both $\kappa$ and $\theta$ contributions of \eqref{torsion-decomposition}). This decomposition rearranges the dependencies of the action on the field variables as
\begin{equation}
S = S [\phi, g_{\mu\nu}, \check{\tau}_\mu, \kappa^\rho{}_{\nu\mu}] \, .
\end{equation}
Since $\delta^{p_2}_v \tau_\mu = \delta^{p_2}_v \check{\tau}_\mu = - (d-1) v_\mu$, we have
\begin{equation}
\delta^{p_2}_v \check{K}^\rho{}_{\nu\mu} = \delta^\rho{}_\mu v_\nu - g_{\nu\mu} v^\rho \, .
\end{equation}
Thus, the second projective variation of the full action yields
\begin{equation}
\begin{split}
 \delta^{p_2}_v S[\phi, g_{\mu\nu}, \check{\tau}_\mu, \kappa^\rho{}_{\nu\mu}]
 &= \int \sqrt{-g} \Sigma_\rho{}^{\nu\mu} \delta^{p_2}_v \check{K}^\rho{}_{\nu\mu}
 \\
 &= 2 \int \sqrt{-g} \Sigma_\rho{}^{\mu\rho} v_\mu \, .
\end{split}
\end{equation}
Again, from the fact that $v_\mu$ is arbitrary we derive the N\"other identity associated with the second projective transformation, $\Sigma_\rho{}^{\mu\rho} = 0$,
which states that the only nontrivial trace of the spin-current has to vanish.

As a consequence of the N\"other identities associated to the first two projective transformations we have derived the condition that the first functional derivatives of the action functional w.r.t.~$S_\mu$ and $\check{\tau}_\mu$ must vanish. That is to say, the action does not depend on either of these two field variables.
Notice that for the derivation of the two N\"other identities it was not necessary to require that the matter field $\phi$ is on-shell.

At this stage we need to remark a somehow subtle point. One might wonder why we bothered derive the N\"other identities to state that the full action has to be independent of $S_\mu$ and $\check{\tau}_\mu$, given that we have previously expressed the same requirements as ``gauge'' choices for the parameters. Indeed, in a given coordinate chart we may cancel them out identically choosing as parameters of the two projective transformations $\xi_\mu = - S_\mu$ and $v_\mu = \frac{1}{d-1} \check{\tau}_\mu$. However, such choice cannot be made globally on the manifold, because $S_\mu$ and $\check{\tau}_\mu$ transform affinely under the Weyl group, whereas the projective parameters are Weyl-invariant. Therefore, we \emph{should} rely on the machinery of N\"other identities to prove that the dependencies of a projective-invariant action functional involving the given field $\phi$ are
\begin{equation}
S = S [\phi, g_{\mu\nu}, \kappa^\rho{}_{\nu\mu}] \, .
\end{equation}
The N\"other identities ensure that the above form holds in every chart.

\section{Example: coupling of fermions in $d=4$}

Now we take into account the possible nonminimal couplings of a Dirac fermion $\psi$ with Riemann-Cartan-Weyl geometry, then we impose projective invariance on all possible actions which are $SL(2,\mathbb{C})\times D_1$ gauge invariant.

First of all, let us summarize some basic properties of spinors in curved spacetime and fix some conventions. The Clifford algebra of Dirac matrices $\gamma^a$ is defined on the local frame satisfying $\{ \gamma^a, \gamma^b \} = 2 \eta^{ab} $.
The covariant derivative of a Dirac spinor is defined as
\begin{equation}
 D_\mu \psi = \partial_\mu \psi + \frac{i}{2} \sigma_{ab} \, \omega^{ab}{}_\mu \psi \, ,
\end{equation}
where $\sigma_{ab}= -\frac{i}{4} [\gamma_a,\gamma_b]$ are the generators of the Lorentz group in the Dirac representation $(\frac{1}{2},0)\oplus(0,\frac{1}{2})$.\footnote{We work with the mostly-plus signature and use the same conventions as Weinberg \cite{Weinberg:1995mt}, see section $5.4$.} The covariant derivative of a gamma matrix vanishes, $ D_\mu \gamma^a = 0$ when appropriately extended to all indices, implying the compatibility of the spin-structure. The hermicity properties of the gamma matrices are encoded by the relation
$ (\gamma^a)^\dagger = \gamma^0 \gamma^a \gamma^0 $.
The pseudo-scalar matrix $\gamma_5$ is defined as
$ \gamma_5 \equiv -i \gamma^0 \gamma^1 \gamma^2 \gamma^3 $, so it is Hermitean
and such that $(\gamma_5)^2 = \mathds{1}$.

Now we have all the necessary ingredients to take into account all the possible ways of coupling a Dirac fermion $\psi$ of Weyl weight $-\frac{3}{2}$ with gravity (in arbitrary dimension it has weight $\frac{1-d}{2}$). Any fermion bi-linear has Weyl weight $w(\overline{\psi}\psi)=-3$, consequently we can saturate the required total Weyl weight of a Lagrangian term with either a derivative, a torsion tensor or a scalar field. Taking into account the irreps of the torsion tensor, we write the allowed interactions
\begin{equation}\label{Dirac-actions}
\begin{split}
  &-\frac{i}{2} \Bigl( \overline{\psi} \gamma^a D_a \psi - \overline{\psi} \overleftarrow{D_a} \gamma^a \psi \Bigr) \, ,
  \quad
  \overline{\psi} \gamma^a \psi \, \tau_a \, ,
  \\&\quad
  \overline{\psi} \gamma^a \gamma_5 \psi \, \theta_a \, ,
  \quad
  \varphi\, \overline{\psi} \psi \, .
\end{split}
\end{equation}
A known fact is that the Weyl potential does not couple to the Dirac Lagrangian
\cite{Gasperini:2017ggf}, which is essentially the first term. This is related to the fact that it is already Weyl invariant. Moreover, the second projective transformation of the Dirac Lagrangian also identically vanishes,
so the nonminimal coupling of the fermionic vector current to the torsion vector given in the second term cannot appear. On the other hand, the axial projective transformation of the Dirac Lagrangian is nonzero
\begin{equation}
 \delta^{p_3}_w \Bigl( \overline{\psi} \gamma^a D_a \psi - \overline{\psi} \overleftarrow{D_a} \gamma^a \psi \Bigr) =  \frac{3}{2} \overline{\psi} \gamma^a \gamma_5 \psi w_a \, .
\end{equation}
Thus we find that the axial projective invariance is satisfied if and only if the third term of \eqref{Dirac-actions} is fixed accordingly to balance the Dirac transformation.
Finally, the Yukawa interaction is not fixed by any requirement.
In conclusion, the most general  projective-invariant action of a Dirac field coupled to a Riemann-Cartan-Weyl geometry is
\begin{equation}
\begin{split}
 \mathcal{L}_{\psi} &= -\frac{i}{2} \left( \overline{\psi} \gamma^a D_a \psi - \overline{\psi} \overleftarrow{D_a} \gamma^a \psi \right)
 \\&
 + \frac{i}{16} \overline{\psi} \gamma^a \gamma_5 \psi \, \theta_a + g \varphi\, \overline{\psi} \psi \, .
\end{split}
\end{equation}

\section{Conclusions}\label{sect:conclusions}

The analysis of the set of transformations of a connection that preserve the projective structure given by the solutions of the autoparallel equation is rich and gives tentalizing new avenues to constrain the proliferation of couplings often seen
in metric-affine theories of gravity.
To a large extent, this rich structure survives when specializing metric-affine theories to Riemann-Cartan-Weyl models in which local dilatations are gauged together with the Lorentz group.

Our most important result is that we have shown how a Riemann-Cartan-Weyl geometry can be constrained, by means of a combination of projective symmetries, to
a model in which the role of the Abelian Weyl gauge potential in played by the vector component of the torsion tensor, referred to as torsion-gauging, while the axial and truly tensorial parts of the torsion are left unchanged. Given that models of gravity with torsion-gauged gauged Weyl symmetry have receveid renewed interest recently,
our constructions and result offer an important framework to discuss the underlying geometry.

The gauging of Weyl symmetry is known to result in models that are actually invariant under (rigid) scale transformations, because the trace of the energy-momentum tensor is a total divergence, rather than zero as it would be for a conformal theory.
Scale invariant models are important tools for more phenomenological applications.
Based on our construction, that is, assuming that the torsion-gauged theory is a projectively-fixed Riemann-Cartan-Weyl model of gravitational interactions,
we deduce that these phenomenological models can, and maybe should, be consistently coupled with the two other irreducible components of the torsion tensor.
These are the axial and truly tensorial part of the torsion (the latter being a Curtright-like tensor with mixed symmetries). We believe that the phenomenological implications of the coupling of these two tensors under the constraint of the projective symmetries is worth studying in the future.

\appendix

\section{Lorentz and Weyl subgroups of $GL(d)$}\label{sect:gld}

The affine algebra of $GL(d)$ is
\begin{equation}
 \begin{split}
 \bigl[G^a{}_b,G^c{}_d \bigr] &= i(\delta^a{}_d G^c{}_b-\delta^c{}_b G^a{}_d) \\
 \bigl[G^a{}_b,P_c \bigr] &= i \delta^a_c P_b \\
 \bigl[P_a,P_b \bigr] &= 0
 \end{split}
\end{equation}
Using the metric $\eta_{ab}$ with the given signature, we can define $G_{ab}=\eta_{ac}G^c{}_b$, that consequently depends on the adopted signature, but the same general idea holds for any other signature. In this way, we can
separate the symmetric and antisymmetric parts of $G_{ab}$, defined $M_{ab}=2G_{(ab)}$ and $J_{ab}=2G_{[ab]}$, respectively.
It is straightforward to check that the subalgebra generated by $J_{ab}$ and $P_a$
is the affine algebra of the Lorentz group $SO(1,d-1)$, that is, the Poincar{\'e} group
of the isometries of $\eta_{ab}$. General Relativity can be seen as emerging from the gauging of the local Lorentz group through the spin-connection $\omega^{a}{}_{b\mu}$.
The same general idea holds for any other signature \cite{Tomboulis:2011qh}.

The affine Lorentz group is a natural subgroup of the affine $GL(d)$, once the signature of the metric is chosen. However, it is possible to find a larger subalgebra, which, besides $J_{ab}$ and $P_a$ contains the trace of the symmetric generators $\eta^{ab}M_{ab}= M^a{}_a$.
In general, we separate the symmetric generators as
$\hat{D}=G^a{}_a=\frac{1}{2}\eta^{ab}M_{ab}$ and $\hat{M}_{ab} = M_{ab}-\frac{2}{d}\eta_{ab} \hat{D}$.
The generator $D$ generates the dilatations, as should evident, for example, from the commutators
\begin{eqnarray}\label{eq:commutators-dilatations}
 \bigl[\hat{D},P_a \bigr] = i P_a \,, \qquad \bigl[\hat{D},J_{ab} \bigr]=0\,,
\end{eqnarray}
which reproduces the expected algebra as can be checked easily from the flat space differential representation $P_a=-i \partial_a$, $J_{ab}=2ix_{[a}\partial_{b]}$ and $\hat{D}=-x^a\partial_a$.
Similarly to the previous case, the Weylian formulation of gravity can be seen
as a gauge theory of the semidirect product of dilatations and Lorentz generators.
The connection of the Weylian formulation generalizes the spin-connection as $\omega^{a}{}_{b\mu}+\delta^a_b S_\mu$.

In the next Appendix we put to use some of the considerations given here on the algebra of generators, but, before moving on, we would like to speculate more on the Goldstone interpretation of the metric \cite{Tomboulis:2011qh,Lindwasser:2022nfa}. We choose $d=4$ for the sake of clarity.
The problem of writing down an explicit $GL(4)$ action that spontaneously breaks down the symmetry to the Lorentz subgroup $SO(1,3)$ has a long story, but no accepted ``definitive'' solution,
although interesting work has been done recently to establish the broken theory $GL(4)\to SO(1,3)$ assuming that fields belong to infinite-dimensional affine spinor representations of $GL(4)$ (i.e., ``affinors'', see Ref.~\cite{Lindwasser:2022nfa} and references therein). There is the possibility of ``intermediate'' breakings in between the two groups, in fact, the $GL(4)$ representations of the proposal of Ref.~\cite{Lindwasser:2022nfa} carry $SL(4)$ indices, so the same breaking pattern could probably be realized by breaking $\hat{D}$ first and $\hat{M}$ later, resulting in $GL(4)\overset{\scriptscriptstyle \hat{D}}{\to} SL(4) \overset{\scriptscriptstyle \hat{M}}{\to} SO(1,3)$. An alternative pattern, we daresay equally amusing, would be to break $\hat{M}$ first and $\hat{D}$ later,
$GL(4) \overset{\scriptscriptstyle \hat{M}}{\to} SO(1,3) \times D_1 \overset{\scriptscriptstyle \hat{D}}{\to} SO(1,3)$,
which is more in line with the considerations of this paper and of the next Appendix ($D_1 \simeq \mathbb{R}$ is the Abelian group of dilatations). The difference between the two patterns lies on when the ``scale'' of gravity is actually generated,
assuming that $GL(4)$ is the ultraviolet symmetry of some quantum theory of gravity
and that the breaking of $\hat{D}$ introduces the Planck mass as effective mass scale.

\section{Coset approach}\label{sect:coset}

We want to offer a relatively brief comparison between some of the geometrical structures that we presented in this paper and the outcome of the coset construction presented in Ref.~\cite{Karananas:2015eha}.
This Appendix is not self-contained, because we just summarize some of the results of Ref.~\cite{Karananas:2015eha} and present them in our notation for an easier comparison. It does allow us to make contact with most of the structures that we have touched upon in the main text and give them a different perspective.

Using the coset construction it is possible to systematically write down the admissible terms of the effective theory that emerges from the spontaneous breaking of one group into another by analyzing the Maurer-Cartan form \cite{Volkov:1973vd}. For this comparison we have in mind
the breaking of $G=GL(d)$ down to the Lorentz subgroup times the Abelian factor generated by the dilatations, denoted here together as $H= SO(1,d-1)\times D_1$.
The degrees of freedom of the effective theory live in the coset $G/H$, to which the Maurer-Cartan form belongs. The resulting theory is manifestly invariant under $H$, but also invariant under $G$, with transformations realized nonlinearly. As customary also in gauged nonlinear sigma-models, the group at the denominator can be gauged \cite{Percacci:1998ag},
so the coset can produce a gauged theory of both Lorentz transformations and dilatations.

Choosing a representative of the coset
which includes the generator of the translations as it is convenient when dealing with spacetime symmetries \cite{Volkov:1973vd},
the authors of Ref.~\cite{Karananas:2015eha} show that the gauged Maurer-Cartan form can be written as
\begin{equation}
 \Omega^{-1} \tilde{\nabla}_\mu \Omega = i e^a{}_\mu P_a +\frac{i}{2} \tilde{\omega}^{ab}{}_\mu J_{ab} + i S_\mu \hat{D} \,,
\end{equation}
where the generators $P_a$, $J_{ab}$ and $\hat{D}$ have been introduced in Appendix \ref{sect:gld}. The effective action for the broken theory then must depend on the curvature of the gauge fields $e^a{}_\mu$, $\tilde{\omega}^{ab}{}_\mu$ and $S_\mu$, in order to be covariant, and can couple to external fields $\psi$ through the covariant derivative
\begin{equation}\label{eq:general-covd-coset}
\begin{split}
 \tilde{\nabla}_a \psi &= E^\mu{}_a \tilde{\nabla}_\mu \psi
 \\
 &=
 E^\mu{}_a \bigl(\partial_\mu +\frac{i}{2} \tilde{\omega}^{ab}{}_\mu J_{ab} + i S_\mu \hat{D} \bigr) \psi\,,
\end{split}
\end{equation}
where the action of the generator of the dilatations on the field is diagonalized by the weight, $\hat{D}\psi=w_\psi \psi$.
Given that the gauge potentials are all independent, this is essentially the same connection that some of us have discussed as starting point in a previous paper \cite{Sauro:2022chz},
but notice that now
we use the symbol $\tilde{\omega}^{ab}{}_\mu$ to differentiate it from
another spin-connection that we are going to encounter soon.

The curvatures associated with the gauge fields are
\begin{equation}
\begin{split}
 (d_{\tilde{\nabla}} e^a{})_{\mu\nu}
 &= 2\partial_{[\mu} e^a{}_{\nu]}
 +2\tilde{\omega}^a{}_{b[\mu} e^b{}_{\nu]}
 +2 S_{[\mu} e^a{}_{\nu]}  \\
 \tilde{R}^{a}{}_{b\mu\nu}
 &=
 2\partial_{[\mu} \tilde{\omega}^a{}_{|b|\nu]}
 + 2\tilde{\omega}^a{}_{c[\mu} \tilde{\omega}^c{}_{|b|\nu]} \\
 W_{\mu\nu} &= \partial_\mu S_\nu - \partial_\nu S_\mu\,,
\end{split}
\end{equation}
and the general effective action for the breaking must depend on scalars formed through them. One important property of the coset construction is that sometimes it is possible to solve algebraically for $H$-covariant quantities to zero, and the symmetry of the resulting effective action is unaffected. The procedure goes under the name of \emph{inverse Higgs mechanism} \cite{Ivanov:1975zq}.

The first possibility, discussed in Ref.~\cite{Karananas:2015eha}, is less important for the main theme of this paper, but we review it for completeness. It involves solving the constraint for a torsionless theory, $(d_{\tilde{\nabla}} e^a{})_{\mu\nu}=0$. This can be done requiring $\tilde{\omega}^{ab}{}_{\mu}= \mathring{\omega}^{ab}{}_{\mu}+(e^a{}_\mu e^b{}_\nu-e^a{}_\nu e^b{}_\mu)S^\nu$.
Under this constraint we have that \eqref{eq:general-covd-coset} becomes
\begin{equation}\label{eq:hat-covd-coset}
\begin{split}
 \tilde{\nabla}_a \psi \to
 \hat{\nabla}_a \psi
  &=
  E^\mu{}_a \bigl(\partial_\mu +\frac{i}{2} \mathring{\omega}^{ab}{}_\mu J_{ab}
  \\&
  +i e^a{}_\mu e^b{}_\nu S^\nu J_{ab} + i S_\mu \hat{D} \bigr) \psi\,,
\end{split}
\end{equation}
which is the torsionless Weyl-covariant derivative discussed in Eq.~\eqref{eq:weyl-gauged-connection} of the main text. The solution of the inverse Higgs constraint produces the additional term that ``mixes'' spacetime and dilatation symmetries.
The construction of Ref.~\cite{Karananas:2015eha} proceeds further by noticing that the tensor $\Theta_{\mu\nu}= \mathring{\nabla}_\mu S_\nu -S_\mu S_\nu + \frac{1}{2} S_\rho S^\rho g_{\mu\nu}$
transforms (independently from $S_\mu$ and) like the Schouten tensor $P_{\mu\nu}=\frac{1}{d-2}\bigl( \mathring{R}_{\mu\nu}-\frac{1}{2(d-1)}g_{\mu\nu}\mathring{R}\bigr)$ under Weyl transformations.

As a consequence, if and only if the effective action depends on $S_\mu$ \emph{only} through the combination $\Theta_{\mu\nu}$, it is possible to replace $\Theta_{\mu\nu} \to P_{\mu\nu}$ everywhere and maintain Weyl invariance. The resulting action will not depend on $S_\mu$ after the replacement, implying that the dilation current is zero, which by itself implies that the resulting theory is invariant under Weyl transformations (conformal) and not only invariant under gauged Weyl transformations. This method, originally developed in \cite{Iorio:1996ad}, goes under the name of \emph{Ricci gauging}. Notice that $\Theta_{\mu\nu}$ should be symmetric under the replacement, implying that we tacitly assume that $W_{\mu\nu}=0$, i.e., the Weyl potential is pure gauge, $S_\mu = \partial_\mu \phi$
for some scalar $\phi$ when Ricci gauging.
A small generalization, briefly noted in Ref.~\cite{Karananas:2015eha} but not fleshed out further, involves the replacement
$\Theta_{\mu\nu} \to P_{\mu\nu} + 2 W_{\mu\nu}$. In this case one has to check that $W_{\mu\nu}$, which still depends on $S_\mu$, decouples from the theory, but, if it does, the Ricci gauging procedure still works.

The second possibility for solving an inverse Higgs constraint is more tightly
connected with the present paper. It comes from noticing that
$(d_{\tilde{\nabla}} e^a{})_{\mu\nu}$ admits three irreducible contributions under the Lorentz decomposition (that is, the decomposition given in \eqref{torsion-decomposition} of the main text). It is thus possible to set the contributions to zero independently. In particular, while the axial and tensor parts do not appear to be useful, the vector one can be solved using $S_\mu$ (as opposed to using $\tilde{\omega}^{ab}{}_{\mu}$ like in the previous case). The inverse Higgs constraint in this case is
$E^\nu{}_a(d_{\tilde{\nabla}} e^a{})_{\mu\nu}=0$, which can be solved algebrically
as $S_\mu=\frac{1}{d-1} \tau_\mu$, where $\tau_\mu$ is the torsion vector
$\tau_\mu=2E^\nu{}_a(\partial_{[\mu} e^a{}_{\nu]}
 +\tilde{\omega}^a{}_{b[\mu} e^b{}_{\nu]})$. The torsion vector takes over the role of the Weyl gauge field, which, as we show in Sect.~\ref{sect:weyl-projective}, is equivalent to requiring the invariance of the Riemann-Cartan-Weyl theory with respect to a combination of projective symmetries. The corresponding covariant derivative is shown in Sect.~\ref{sect:comments}.
Similarly to the first possibility, we could call this procedure \emph{torsion gauging} of the Weyl symmetry.
Notice however that the second choice for the inverse Higgs constraint sets to zero only the vector part of $(d_{\tilde{\nabla}} e^a{})_{\mu\nu}$, implying that the axial and tensor components of the torsion are covariant and should still be present in the resulting theory (unless they are set to zero by some other requirement), which is precisely what we observe after requiring the invariances under the projective symmetries!

\section{Projective transformations of the curvatures}\label{sect:variations}

We list here the projective transformations of the the relevant field-strengths of Riemann-Cartan-Weyl geometry. We start from the only two irreducible components of the torsion which are affected by such transformations. For the torsion vector we have:
\begin{equation}\label{proj-vector-tor}
 \begin{split}
 \delta^{p_1}_\xi \tau_a & =  E_b \lfloor E_a \lfloor (\xi \wedge e^b) = (d-1) \xi_a \, ,\\
 \delta^{p_2}_v \tau_a & = E_b \lfloor E_a \lfloor (e^b \wedge v) = - (d-1) v_a \, ,\\
 \delta^{p_3}_w \tau_a & = 0 \, .
 \end{split}
\end{equation}
On the other hand, the only nontrivial variation of the axial-torsion is the last one,
which we give in $d=4$
\begin{equation}\label{proj-axial-tor}
\begin{split}
\delta^{p_3}_w \theta_a
 &= \varepsilon_{ab}{}^{cd} \, \varepsilon^{gb}{}_{fh} \, w_g \, E_d \lfloor E_c \lfloor ( e^h \wedge e^f)
 \\&
 = - 2 \varepsilon_{ab}{}^{cd} \varepsilon^{gb}{}_{cd} w_g = 12 w_a \, .
\end{split}
\end{equation}
Another useful transformation is that of the covariant derivative of a scalar field $\varphi$ with weight $w(\varphi)=w_\varphi$
\begin{equation}\label{proj1-grad-scalar}
 \delta^{p_1}_\xi \tilde{\nabla}_a \varphi = w_\varphi \xi_a \varphi \, ,
\end{equation}
whereas the other two transformations vanish identically.

Now we turn to the projective transformations of the Riemann and Weyl $2$-forms. Since we are using the geometrical language of differential forms, the Riemann $2$-form is unaffected by $\delta^{p_1}_\xi$, while the Weyl $2$-form is insensitive to $\delta^{p_2}_v$ and $\delta^{p_3}_w$. The variation of the Riemann $2$-form under the second projective transformation reads
\begin{equation}\label{proj2-Riem-2form}
\begin{split}
\delta^{p_2}_v \mathcal{R}^a{}_b &=
 d \delta^{p_2}_v \omega^a{}_b + \delta^{p_2}_v (\omega^a{}_c \wedge \omega^c{}_b)
 \\ &= v_b \, \mathcal{T}^a - v^a \, \mathcal{T}_b + D v_b \wedge e^a - D v^a \wedge e_b
 \\&
 + v_b \, e^a \wedge v - v^a \, e_b \wedge v - v^2 \, e^a \wedge e^b \, .
\end{split}
\end{equation}
The action of the third projective transformation is
\begin{equation}\label{proj3-Riem-2form}
\begin{split}
\delta^{p_3}_w \mathcal{R}^a{}_b &=  \varepsilon^{da}{}_{bc} \, w_d \, \mathcal{T}^c + \varepsilon^{da}{}_{bc} \, D w_d \wedge e^c
\\&
+ w_b \, e^a \wedge w - w^a \, e_b \wedge w - w^2 \, e^a \wedge e_b \, .
\end{split}
\end{equation}
At last, we consider the Weyl $2$-form, also known as homothetic curvature. As mentioned above, its only nontrivial projective transformation is the first one, which reads
\begin{equation}
 \delta^{p_1}_\xi \mathcal{W} = d \xi \, .
\end{equation}



\begin{thebibliography}{10}
\ifx\href\asklfhas\newcommand{\href}[2]{#2}\fi
\ifx\arxivref\asklfhas\newcommand{\arxivref}[2]{\href{http://arxiv.org/abs/#1}{#2}}\fi
\ifx\doiref\asklfhas\newcommand{\doiref}[2]{\href{http://dx.doi.org/#1}{#2}}\fi
\parskip 0pt
\normalsize



\bibitem{Schouten2013ricci}
J.~A. Schouten,
\textit{``Ricci-calculus: an introduction to tensor analysis and its geometrical applications"},
Springer Science \& Business Media (2013)\ignorespaces
\bibitem{kobayashi1964projective}
S.~Kobayashi \& T.~Nagano,
\textit{``On projective connections''},
Journal~of~Mathematics~and~Mechanics \textbf{}, 215
  (1964)\ignorespaces\ignorespaces
\bibitem{ehlers1973geometry}
J.~Ehlers \& A.~Schild,
\textit{``Geometry in a manifold with projective structure''},
Communications~in~Mathematical~Physics \textbf{32}, 119
  (1973)\ignorespaces\ignorespaces
\bibitem{aminova1995projective}
A.~V. Aminova,
\textit{``Projective transformations and symmetries of differential
  equation''},
Sbornik:~Mathematics \textbf{186}, 1711 (1995)\ignorespaces\ignorespaces
\bibitem{giachetta1997projective}
G.~Giachetta \& L.~Mangiarotti,
\textit{``Projective invariance and Einstein's equations''},
General~Relativity~and~Gravitation \textbf{29}, 5
  (1997)\ignorespaces\ignorespaces
\bibitem{matveev2018projectively}
V.~S. Matveev,
\textit{``Projectively invariant objects and the index of the group of affine
  transformations in the group of projective transformations''},
Bulletin~of~the~Iranian~Mathematical~Society \textbf{44}, 341
  (2018)\ignorespaces\ignorespaces
\bibitem{Afonso:2017bxr}
V.~I. Afonso, C.~Bejarano, J.~Beltran~Jimenez, G.~J. Olmo \& E.~Orazi,
\textit{``{The trivial role of torsion in projective invariant theories of
  gravity with non-minimally coupled matter fields}''},
\doiref{10.1088/1361-6382/aa9151}{Class.~Quant.~Grav. \textbf{34}, 235003
  (2017)\ignorespaces}\ignorespaces,
\normalsize{\texttt{\arxivref{1705.03806}{arXiv:1705.03806
  \![gr-qc]}}}\ignorespaces
\bibitem{Aoki:2019rvi}
K.~Aoki \& K.~Shimada,
\textit{``{Scalar-metric-affine theories: Can we get ghost-free theories from
  symmetry?}''},
\doiref{10.1103/PhysRevD.100.044037}{Phys.~Rev.~D \textbf{100}, 044037
  (2019)\ignorespaces}\ignorespaces,
\normalsize{\texttt{\arxivref{1904.10175}{arXiv:1904.10175
  \![hep-th]}}}\ignorespaces
\bibitem{BeltranJimenez:2019acz}
J.~Beltr\'an~Jim\'enez \& A.~Delhom,
\textit{``{Ghosts in metric-affine higher order curvature gravity}''},
\doiref{10.1140/epjc/s10052-019-7149-x}{Eur.~Phys.~J.~C \textbf{79}, 656
  (2019)\ignorespaces}\ignorespaces,
\normalsize{\texttt{\arxivref{1901.08988}{arXiv:1901.08988
  \![gr-qc]}}}\ignorespaces
\bibitem{BeltranJimenez:2020sqf}
J.~Beltr\'an~Jim\'enez \& A.~Delhom,
\textit{``{Instabilities in metric-affine theories of gravity with higher order
  curvature terms}''},
\doiref{10.1140/epjc/s10052-020-8143-z}{Eur.~Phys.~J.~C \textbf{80}, 585
  (2020)\ignorespaces}\ignorespaces,
\normalsize{\texttt{\arxivref{2004.11357}{arXiv:2004.11357
  \![gr-qc]}}}\ignorespaces
\bibitem{BeltranJimenez:2020guo}
J.~Beltr\'an~Jim\'enez, D.~De~Andr\'es \& A.~Delhom,
\textit{``{Anisotropic deformations in a class of projectively-invariant
  metric-affine theories of gravity}''},
\doiref{10.1088/1361-6382/abb923}{Class.~Quant.~Grav. \textbf{37}, 225013
  (2020)\ignorespaces}\ignorespaces,
\normalsize{\texttt{\arxivref{2006.07406}{arXiv:2006.07406
  \![gr-qc]}}}\ignorespaces
\bibitem{Klemm:2020mfp}
S.~Klemm \& L.~Ravera,
\textit{``{An action principle for the Einstein–Weyl equations}"}
\doiref{10.1016/j.geomphys.2020.103958}{J.~Geom.~Phys. \textbf{158}, 103958 (2020) \ignorespaces}\ignorespaces,
\normalsize{\texttt{\arxivref{2006.15890}{arXiv:2006.15890
 \![hep-th]}}}\ignorespaces
\bibitem{Garcia-Parrado:2020lpt}
A.~Garc\'\i{}a-Parrado \& E.~Minguzzi,
\textit{``{Projective and amplified symmetries in metric-affine theories}''},
\doiref{10.1088/1361-6382/abed61}{Class.~Quant.~Grav. \textbf{38}, 135001
	(2021)\ignorespaces}\ignorespaces,
\normalsize{\texttt{\arxivref{2006.04040}{arXiv:2006.04040
			\![gr-qc]}}}\ignorespaces
\bibitem{Iosifidis:2018zwo}
D.~Iosifidis \& T.~Koivisto,
\textit{``{Scale transformations in metric-affine geometry}"}
\doiref{10.3390/universe5030082}{Universe \textbf{5}, 82 (2019) \ignorespaces}\ignorespaces,
\normalsize{\texttt{\arxivref{1810.12276}{arXiv:1810.12276
 \![gr-qc]}}}\ignorespaces
\bibitem{Iosifidis:2018zij}
D.~Iosifidis, A.~C. Petkou \& C.~G. Tsagas,
\textit{``{Torsion/non-metricity duality in f(R) gravity}"}
\doiref{10.1007/s10714-019-2539-9}{Gen.~Rel.~Grav. \textbf{51}, 66 (2019) \ignorespaces}\ignorespaces,
\normalsize{\texttt{\arxivref{1810.06602}{arXiv:1810.06602
 \![gr-qc]}}}\ignorespaces
\bibitem{Percacci:2020ddy}
R.~Percacci \& E.~Sezgin,
\textit{``{New class of ghost- and tachyon-free metric affine gravities}''},
\doiref{10.1103/PhysRevD.101.084040}{Phys.~Rev.~D \textbf{101}, 084040
  (2020)\ignorespaces}\ignorespaces,
\normalsize{\texttt{\arxivref{1912.01023}{arXiv:1912.01023
  \![hep-th]}}}\ignorespaces
\bibitem{Baldazzi:2021kaf}
A.~Baldazzi, O.~Melichev \& R.~Percacci,
\textit{``{Metric-Affine Gravity as an effective field theory}''},
\doiref{10.1016/j.aop.2022.168757}{Annals~Phys. \textbf{438}, 168757
(2022)\ignorespaces}\ignorespaces,
\normalsize{\texttt{\arxivref{2112.10193}{arXiv:2112.10193
\![gr-qc]}}}\ignorespaces
\bibitem{Iosifidis:2019fsh}
D.~Iosifidis,
\textit{``{Linear Transformations on Affine-Connections}"}
\doiref{10.1088/1361-6382/ab778d}{Class.~Quant.~Grav. \textbf{37}, 085010 (2020) \ignorespaces}\ignorespaces,
\normalsize{\texttt{\arxivref{1911.04535}{arXiv:1911.04535
 \![gr-qc]}}}\ignorespaces
\bibitem{Charap:1973fi}
J.~M. Charap \& W.~Tait,
\textit{``{A GAUGE THEORY OF THE WEYL GROUP}''},
\doiref{10.1098/rspa.1974.0151}{Proc.~Roy.~Soc.~Lond.~A \textbf{340}, 249
  (1974)\ignorespaces}\ignorespaces
\bibitem{Gasperini:2017ggf}
M.~Gasperini,
\textit{``{Theory of Gravitational Interactions}''},
Springer International Publishing (2017)\ignorespaces,
Cham
\bibitem{Ghilencea:2018dqd}
D.~M. Ghilencea,
\textit{``{Spontaneous breaking of Weyl quadratic gravity to Einstein action
  and Higgs potential}''},
\doiref{10.1007/JHEP03(2019)049}{JHEP \textbf{1903}, 049
  (2019)\ignorespaces}\ignorespaces,
\normalsize{\texttt{\arxivref{1812.08613}{arXiv:1812.08613
  \![hep-th]}}}\ignorespaces
\bibitem{Ghilencea:2018thl}
D.~M. Ghilencea \& H.~M. Lee,
\textit{``{Weyl gauge symmetry and its spontaneous breaking in the standard
  model and inflation}''},
\doiref{10.1103/PhysRevD.99.115007}{Phys.~Rev.~D \textbf{99}, 115007
  (2019)\ignorespaces}\ignorespaces,
\normalsize{\texttt{\arxivref{1809.09174}{arXiv:1809.09174
  \![hep-th]}}}\ignorespaces
\bibitem{Ghilencea:2019jux}
D.~M. Ghilencea,
\textit{``{Stueckelberg breaking of Weyl conformal geometry and applications to
  gravity}''},
\doiref{10.1103/PhysRevD.101.045010}{Phys.~Rev.~D \textbf{101}, 045010
  (2020)\ignorespaces}\ignorespaces,
\normalsize{\texttt{\arxivref{1904.06596}{arXiv:1904.06596
  \![hep-th]}}}\ignorespaces
\bibitem{Karananas:2021gco}
G.~K. Karananas, M.~Shaposhnikov, A.~Shkerin \& S.~Zell,
\textit{``{Scale and Weyl invariance in Einstein-Cartan gravity}''},
\doiref{10.1103/PhysRevD.104.124014}{Phys.~Rev.~D \textbf{104}, 124014
  (2021)\ignorespaces}\ignorespaces,
\normalsize{\texttt{\arxivref{2108.05897}{arXiv:2108.05897
  \![hep-th]}}}\ignorespaces
\bibitem{Karananas:2015eha}
G.~K. Karananas \& A.~Monin,
\textit{``{Weyl and Ricci gauging from the coset construction}''},
\doiref{10.1103/PhysRevD.93.064013}{Phys.~Rev.~D \textbf{93}, 064013
  (2016)\ignorespaces}\ignorespaces,
\normalsize{\texttt{\arxivref{1510.07589}{arXiv:1510.07589
  \![hep-th]}}}\ignorespaces
\bibitem{Curtright:1980yk}
T.~Curtright,
\textit{``{GENERALIZED GAUGE FIELDS}''},
\doiref{10.1016/0370-2693(85)91235-3}{Phys.~Lett.~B \textbf{165}, 304
  (1985)\ignorespaces}\ignorespaces
\bibitem{Weinberg:1972kfs}
S.~Weinberg,
\textit{``{Gravitation and Cosmology}: {Principles and Applications of the
  General Theory of Relativity}''},
John Wiley and Sons (1972)\ignorespaces,
New York
\bibitem{Wald:1984rg}
R.~M. Wald,
\textit{``{General Relativity}''},
Chicago Univ. Pr. (1984)\ignorespaces,
Chicago, USA
\bibitem{Hehl:1976my}
F.~W. Hehl, G.~D. Kerlick \& P.~Von Der~Heyde,
\textit{``{On a New Metric Affine Theory of Gravitation}''},
\doiref{10.1016/0370-2693(76)90393-2}{Phys.~Lett.~B \textbf{63}, 446
  (1976)\ignorespaces}\ignorespaces
\bibitem{Percacci:1990wy}
R.~Percacci,
\textit{``{The Higgs phenomenon in quantum gravity}''},
\doiref{10.1016/0550-3213(91)90510-5}{Nucl.~Phys.~B \textbf{353}, 271
  (1991)\ignorespaces}\ignorespaces,
\normalsize{\texttt{\arxivref{0712.3545}{arXiv:0712.3545
  \![hep-th]}}}\ignorespaces
\bibitem{Hehl:1994ue}
F.~W. Hehl, J.~D. McCrea, E.~W. Mielke \& Y.~Ne'eman,
\textit{``{Metric affine gauge theory of gravity: Field equations, Noether
  identities, world spinors, and breaking of dilation invariance}''},
\doiref{10.1016/0370-1573(94)00111-F}{Phys.~Rept. \textbf{258}, 1
  (1995)\ignorespaces}\ignorespaces,
\normalsize{\texttt{\arxivref{gr-qc/9402012}{gr-qc/9402012}}}\ignorespaces
\bibitem{Percacci:2009ij}
R.~Percacci,
\textit{``{Gravity from a Particle Physicists' perspective}''},
\doiref{10.22323/1.081.0011}{PoS \textbf{ISFTG}, 011
  (2009)\ignorespaces}\ignorespaces,
\normalsize{\texttt{\arxivref{0910.5167}{arXiv:0910.5167
  \![hep-th]}}}\ignorespaces
\bibitem{BeltranJimenez:2019esp}
J.~Beltr\'an~Jim\'enez, L.~Heisenberg \& T.~S. Koivisto,
\textit{``{The Geometrical Trinity of Gravity}''},
\doiref{10.3390/universe5070173}{Universe \textbf{5}, 173
  (2019)\ignorespaces}\ignorespaces,
\normalsize{\texttt{\arxivref{1903.06830}{arXiv:1903.06830
  \![hep-th]}}}\ignorespaces
\bibitem{Lindwasser:2022nfa}
L.~W. Lindwasser \& E.~T. Tomboulis,
\textit{``{Searching for Gravity Without a Metric}''},
\normalsize{\texttt{\arxivref{2207.01067}{arXiv:2207.01067
  \![hep-th]}}}\ignorespaces
\bibitem{Matveev:2020wif}
V.~S. Matveev \& E.~Scholz,
\textit{``{Light cone and Weyl compatibility of conformal and projective
  structures}''},
\doiref{10.1007/s10714-020-02716-9}{Gen.~Rel.~Grav. \textbf{52}, 66
  (2020)\ignorespaces}\ignorespaces,
\normalsize{\texttt{\arxivref{2001.01494}{arXiv:2001.01494
  \![math.DG]}}}\ignorespaces
\bibitem{thomas:1934differential}
T.~Y. Thomas,
\textit{``{The differential invariants of generalized spaces}"},
The University Press (1934)\ignorespaces
\bibitem{Iorio:1996ad}
A.~Iorio, L.~O'Raifeartaigh, I.~Sachs \& C.~Wiesendanger,
\textit{``{Weyl gauging and conformal invariance}''},
\doiref{10.1016/S0550-3213(97)00190-9}{Nucl.~Phys.~B \textbf{495}, 433
  (1997)\ignorespaces}\ignorespaces,
\normalsize{\texttt{\arxivref{hep-th/9607110}{hep-th/9607110}}}\ignorespaces
\bibitem{Sauro:2022chz}
D.~Sauro \& O.~Zanusso,
\textit{``{The origin of Weyl gauging in metric-affine theories}''},
\doiref{10.1088/1361-6382/ac82a2}{Class.~Quant.~Grav. \textbf{39}, 185001
  (2022)\ignorespaces}\ignorespaces,
\normalsize{\texttt{\arxivref{2203.08692}{arXiv:2203.08692
  \![hep-th]}}}\ignorespaces
\bibitem{Wheeler:2022ggm}
J.~T. Wheeler,
\textit{``{Abelian symmetry and the Palatini variation}''},
\normalsize{\texttt{\arxivref{2201.00938}{arXiv:2201.00938
  \![gr-qc]}}}\ignorespaces
\bibitem{Scholz:2011za}
E.~Scholz,
\textit{``{Weyl geometry in late 20th century physics}''},
\normalsize{\texttt{\arxivref{1111.3220}{arXiv:1111.3220
  \![math.HO]}}}\ignorespaces
\bibitem{Weinberg:1995mt}
S.~Weinberg,
\textit{``{The Quantum theory of fields. Vol. 1:}: {Foundations}''},
Cambridge University Press (2005)\ignorespaces
\bibitem{Tomboulis:2011qh}
E.~T. Tomboulis,
\textit{``{General Relativity as the effective theory of GL(4,R) spontaneous
  symmetry breaking}''},
\doiref{10.1103/PhysRevD.84.084018}{Phys.~Rev.~D \textbf{84}, 084018
  (2011)\ignorespaces}\ignorespaces,
\normalsize{\texttt{\arxivref{1105.5848}{arXiv:1105.5848
  \![hep-th]}}}\ignorespaces
\bibitem{Volkov:1973vd}
D.~V. Volkov,
\textit{``{Phenomenological Lagrangians}''},
Fiz.~Elem.~Chast.~Atom.~Yadra \textbf{4}, 3 (1973)\ignorespaces\ignorespaces
\bibitem{Percacci:1998ag}
R.~Percacci \& E.~Sezgin,
\textit{``{Properties of gauged sigma models}''},
in \textit{``{Richard Arnowitt Fest: A Symposium on Supersymmetry and
  Gravitation}''},
255--278\ignorespaces\bibitem{Ivanov:1975zq}
E.~A. Ivanov \& V.~I. Ogievetsky,
\textit{``{The Inverse Higgs Phenomenon in Nonlinear Realizations}''},
\doiref{10.1007/BF01028947}{Teor.~Mat.~Fiz. \textbf{25}, 164
  (1975)\ignorespaces}\ignorespaces
\end{thebibliography}

\end{document}